\documentclass[twocolumn,showpacs,prb]{revtex4-1}%
\usepackage{amsfonts}
\usepackage{amsmath}
\usepackage{amssymb}
\usepackage{graphicx}
\usepackage{txfonts}%
\providecommand{\U}[1]{\protect\rule{.1in}{.1in}}

\begin{document}
\title{Theory of nuclear spin dephasing and relaxation by optically illuminated
nitrogen-vancy center}
\author{Wang Ping}
\affiliation{Hefei National Laboratory for Physics Sciences at Microscale and Department of
Modern Physics, University of Science and Technology of China, Hefei, Anhui
230026, China}
\affiliation{Beijing Computational Science Research Center, Beijing 100084, China}
\author{Wen Yang}
\email{wenyang@csrc.ac.cn}
\affiliation{Beijing Computational Science Research Center, Beijing 100084, China}

\begin{abstract}
Dephasing and relaxation of the nuclear spins coupled to the nitrogen-vacancy
(NV) center during optical initialization and readout is an important issue
for various applications of this hybrid quantum register. Here we present both
an analytical description and a numerical simulation for this process, which
agree reasonably with the experimental measurements. For the NV center under
cyclic optical transition, our analytical formula not only provide a clear
physics picture, but also allows controlling the nuclear spin dissipation by
tuning an external magnetic field. For more general optical pumping,
our analytical formula reveals significant contribution to the nuclear spin
dissipation due to electron random hopping into/out of the $m=0$ (or $m=\pm1$)
subspace. This contribution is not suppressed even under saturated optical
pumping and/or vanishing magnetic field, thus providing a possible solution to
the puzzling observation of nuclear spin dephasing in zero perpendicular
magnetic field [M. V. G. Dutt \textit{et al}., Science \textbf{316}, 1312
(2007)]. It also implies that enhancing the degree of spin polarization of the
nitrogen-vacancy center can reduce the effect of optical induced nuclear spin dissipation.

\end{abstract}

\maketitle

\section{Introduction}

Diamond nitrogen-vacancy (NV) center \cite{GruberScience1997} is a leading
platform for various quantum technologies such as quantum communication,
quantum computation, and nanoscale sensing
\cite{DuttScience2007,MazeNature2008,DoldeNatPhys2011,ChildressScience2006,ToganNature2010,NeumannNatPhys2010,NeumannScience2010}%
. The electronic spin of the NV center and a few surrounding nuclear spins
form a hybrid quantum register
\cite{NeumannScience2008,YaoNatCommun2012,WaldherrNature2014}. Important
advantages of this solid-state quantum register include the long electron and
nuclear spin coherence time \cite{MaurerScience2012}, the capability of
high-fidelity initialization, coherent manipulation, and projective readout of
the electronic/nuclear spins \cite{JelezkoPRL2004a,ChildressScience2006} and
even the entire quantum register
\cite{PfaffNatPhys2013,TaminiauNatNano2014,WaldherrNature2014} by optical and
microwave (or radio frequency) illumination. However, during the optical
illumination for initialization and readout
\cite{DuttScience2007,JiangPRL2008,NeumannScience2010,DreauPRL2013,PfaffNatPhys2013,BlokNatPhys2014}%
, the dissipative spontaneous emission and non-radiative decay of the NV
electron generates substantial noise on the nuclear spin qubits through the
hyperfine interaction (HFI), which may significantly degrade the control
precesion. This motivates widespread interest in using the NV center electron
to engineer the nuclear spin dissipation, including pure dephasing and
relaxation \cite{JiangPRL2008,DreauPRL2013}.

In the past few years, the optically induced nuclear spin dissipation has been
investigated in many works
\cite{JacquesPRL2009,FischerPRB2013,WangNatCommun2013,JiangPRL2008,DreauPRL2013}%
. Generally, the nuclear spin dissipation originates from the random
fluctuation of the NV electron under optical illumination, which falls into
two categories: one involving the flip of the NV electron spin and the other
does not. The former is usually strongly suppressed by the large energy
splitting of the NV electron unless the NV electron is tuned to the ground
state or excited state level anticrossing
\cite{JacquesPRL2009,WangNatCommun2013}. The latter is energetically more
favorable and dominates the nuclear spin dissipation in many situations, as
confirmed by a series of experiments
\cite{DreauPRL2013,MaurerScience2012,DuttScience2007}. The theoretical
investigation of this latter mechanism has been carried out in the framework
of a phenomenological spin-fluctuator model \cite{JiangPRL2008}. This work
gives an intuitive understanding for the optically induced nuclear spin
dissipation: the generation of a rapidly fluctuating effective magnetic field
on the nuclear spins by the optically induced random hopping of the electron
between different states. When the hopping is sufficiently fast and hence the
noise correlation time is sufficiently short, the nuclear spin dissipation
could be suppressed \cite{JiangPRL2008} in a way similar to the motional
narrowing effect in NMR spectroscopy in liquids. This effect has been
successfully used to significantly increase the nuclear spin coherence time
\cite{MaurerScience2012}.

Despite these remarkable success, this spin-fluctuator model still suffers
from two drawbacks. First, its analytical form is qualitative, while obtaining
quantitative results require numerical simulations. This not only complicates
the calculation, but also smears the underlying physics picture. Second, the
various parameters in this model are phenomenological, i.e., they are not
directly related to the physical parameters of the NV center, but instead are
obtained from fitting the experimental data. This precludes a straightforward
guidance on controlling the nuclear spin dissipation by tuning various
experimental parameters.

To bridge this gap between experimental observation and theoretical
understanding, we present a microscopic and analytical theory on the nuclear
spin dephasing and relaxation by an optically illuminated NV center at room
temperature. In addition to performing numerical simulation of the coupled
NV-nuclear spin evolution, we further derive analytically a closed Lindblad
master equation for the nuclear spin by adiabatically eliminating the fast
electron spin dynamics in the Born-Markovian approximation. We begin with the
simplest case in which a single cyclic transitions (e.g., between the ground
and excited $m=0$ states) of the NV center is optically driven. Our analytical
expressions for the nuclear spin dephasing and relaxation provide a
quantitative description and a physically transparent interpretation that
substantiates the previous analytical (but qualitative) and numerical results
\cite{JiangPRL2008}. They also demonstrate the possibility to control the
nuclear spin dissipation by tuning the magnetic field \cite{JiangPRL2008}.
Next we consider general optical illumination of the NV center incorporating
finite inter-system crossing between $m=0$ and $m=\pm1$ subspaces. Our
numerical results agree well with the experimental measurements
\cite{DreauPRL2013}. Our analytical results shows that the random hopping
between the $m=0$ (or $m=\pm1$) triplet states and the metastable singlet of
the NV center could significantly contribute to nuclear spin dissipation. This
contribution is not suppressed under saturated optical pumping and is nearly
independent of the magnetic field. This provides a possible solution to the
puzzling observation of nuclear spin dephasing in zero magnetic field
\cite{DuttScience2007}. An analytical formula for the nuclear spin dissipation
in terms of the HFI tensors also allows us to measure the HFI tensor for the
excited electron state, which is usually smeared by the short electron
spontaneous emission lifetime.

\section{Two-level fluctuator model: analytical results}

\subsection{Model}

To begin with, we present a microscopic theory for the decoherence of an
arbitrary nuclear spin $\hat{\mathbf{I}}$ (e.g., $^{13}$C, $^{15}$N, or
$^{14}$N) by the electron of the NV center undergoing optically induced cyclic
transition $|g\rangle\leftrightarrow|e\rangle$, e.g., $|g\rangle=|0\rangle$
and $|e\rangle=|E_{y}\rangle$ in the widely used setup for single-shot readout
\cite{RobledoNature2011,PfaffNatPhys2013}. In the rotating frame, the electron
dynamics is governed by the Liouville superoperator $\mathcal{L}_{e}%
(\cdot)\equiv-i[\hat{H}_{e},(\cdot)]+\sum_{\alpha}\gamma_{\alpha}%
\mathcal{D}(\hat{L}_{\alpha})(\cdot)$, where
\[
\hat{H}_{e}=\Delta\hat{\sigma}_{e,e}+\frac{\Omega_{R}}{2}(\hat{\sigma}%
_{e,g}+h.c)
\]
is the electron Hamiltonian, $\hat{\sigma}_{i,j}\equiv|i\rangle\langle j|$,
$\Delta$ is the detuning of the optical pumping, and $\gamma_{\alpha}$ is the
rate of the $\alpha$th dissipation process $\hat{L}_{\alpha}$ in the Lindblad
form $\mathcal{D}(\hat{L}_{\alpha})(\cdot)\equiv\hat{L}_{\alpha}(\cdot)\hat
{L}_{\alpha}^{\dagger}-\{\hat{L}_{\alpha}^{\dagger}\hat{L}_{\alpha}%
,(\cdot)\}/2$. Here we include the spontaneous emission $\hat{L}=\hat{\sigma
}_{g,e}$ from $|e\rangle$ to $|g\rangle$ with rate $\gamma_{1}\approx
1/(12\ \mathrm{ns})$ and the pure dephasing $\hat{L}=\hat{\sigma}_{e,e}$ of
the excited state $|e\rangle$ with rate $\gamma_{\varphi}$, which has a strong
temperature dependence, from a few tens of $\mathrm{MHz}$ at low temperature
up to $10^{7}\ \mathrm{MHz}$ at room temperature \cite{AbtewPRL2011,FuPRL2009}%
. Including the electron-nuclear HFI $(\hat{\mathbf{S}}_{g}\cdot\mathbf{A}%
_{g}+\hat{\mathbf{S}}_{e}\cdot\mathbf{A}_{e})\cdot\hat{\mathbf{I}}\equiv
\hat{\mathbf{F}}\cdot\hat{\mathbf{I}}$ and the nuclear spin Zeeman term
$\gamma_{N}{\mathbf{B}}\cdot\hat{\mathbf{I}}$ ($\gamma_{N}=-10.705$
\textrm{kHz/mT} is the $^{13}$C nuclear gyromagnetic ratio) under a magnetic
field $\mathbf{B}$, the electron-nuclear coupled system obeys%
\begin{equation}
\dot{\rho}=\mathcal{L}_{e}\hat{\rho}-i[(\hat{\mathbf{F}}+\gamma_{N}%
{\mathbf{B}})\cdot\hat{\mathbf{I}},\hat{\rho}] \label{ME0}%
\end{equation}
in the rotating frame of the pumping laser.

There are two contributions to the nuclear spin dissipation. One involves the
flip of the electron spin and hence is strongly suppressed by the large
electron-nuclear energy mismatch away from the NV center ground state and
excited state anticrossings. The other does not flip the electron spin and
hence is energetically favorable in most situations. In our analytical
derivation, we neglect the former contribution by dropping the off-diagonal
electron spin flip terms in $\hat{\mathbf{F}}$ and only keep the diagonal
part:\ $\hat{\mathbf{F}}\approx\hat{\sigma}_{g,g}\boldsymbol{\omega}_{g}%
+\hat{\sigma}_{e,e}\boldsymbol{\omega}_{e}$, where $\boldsymbol{\omega}%
_{g}=\langle g|\hat{\mathbf{S}}_{g}|g\rangle\cdot\mathbf{A}_{g}$ and
$\boldsymbol{\omega}_{e}=\langle e|\hat{\mathbf{S}}_{e}|e\rangle
\cdot\mathbf{A}_{e}$. The second term of Eq. (\ref{ME0}) describes the
precession of the nuclear spin with angular frequency $\gamma_{N}{\mathbf{B}%
}+\boldsymbol{\omega}_{g}$ and $\gamma_{N}{\mathbf{B}}+\boldsymbol{\omega}%
_{e}$, respectively, conditioned on the electron state being $|g\rangle$ and
$|e\rangle$. When $\boldsymbol{\omega}_{g}\neq\boldsymbol{\omega}_{e}$, the
optically induced random hopping of the electron between $|g\rangle$ and
$|e\rangle$ gives rise to random fluctuation of the nuclear spin precession
frequency and hence nuclear spin dissipation: the fluctuation of the
precession frequency orientation (magnitude)\ leads to nuclear spin relaxation
(pure dephasing) \cite{JiangPRL2008}. Below we derive analytical a closed
equation of motion of the nuclear spin to describe these effects.

\subsection{Lindblad master equation for nuclear spin}

The time scale for the optically pumped two-level NV center to reach its
steady state is $\sim\tau_{\mathrm{NV}}\equiv1/(2R+\gamma_{1})<12\ \mathrm{ns}%
$, where $R=2\pi(\Omega_{R}/2)^{2}\delta^{((\gamma_{1}+\gamma_{\varphi}%
)/2)}(\Delta)$ is the optical pumping rate from $|g\rangle$ to $|e\rangle$ and
$\delta^{(\gamma)}(x)=(\gamma/\pi)/(x^{2}+\gamma^{2})$ is the broadened
$\delta$-function. When $\tau_{\mathrm{NV}}$ is much shorter than the time
scale of the nuclear spin dissipation, we can regard the NV center as always
in its steady state $\hat{P}$ as determined by $\mathcal{L}_{e}\hat{P}=0$,
e.g., the steady state population on $|e\rangle$ and $|g\rangle$ are
$P_{e}=R/(2R+\gamma_{1})$ and $P_{g}=1-P_{e}$, respectively. Then we treat the
dissipative NV center as a Markovian bath \cite{WangArxiv2015} and use
Born-Markovian approximation to derive a Lindblad master equation for the
reduced density matrix of the nuclear spin $\hat{p}(t)\equiv\operatorname*{Tr}%
_{e}\hat{\rho}(t)$ (see appendix A for details):
\begin{equation}
\dot{p}=-i[\boldsymbol{\bar{\omega}}\cdot\hat{\mathbf{I}},\hat{p}%
]+2\Gamma_{\varphi}\mathcal{D}[\hat{I}_{Z}]\hat{p}+\Gamma_{+}\mathcal{D}%
[\hat{I}_{+}]\hat{p}+\Gamma_{-}\mathcal{D}[\hat{I}_{-}]\hat{p}, \label{ME_N}%
\end{equation}
where $\boldsymbol{\bar{\omega}}\equiv\gamma_{N}\mathbf{B}+P_{g}%
\boldsymbol{\omega}_{g}+P_{e}\boldsymbol{\omega}_{e}$ is the average
precession frequency that defines the nuclear spin quantization axis
$\mathbf{e}_{Z}\equiv\boldsymbol{\bar{\omega}}/|\boldsymbol{\bar{\omega}}|$.
The last three terms describe the nuclear spin dissipation in the tilted
cartesian coordinate
\begin{subequations}
\label{XYZ}%
\begin{align}
\mathbf{e}_{X}  &  =\mathbf{e}_{x}\sin\varphi-\mathbf{e}_{y}\cos\varphi,\\
\mathbf{e}_{Y}  &  =\cos\varphi\cos\theta\mathbf{e}_{x}+\sin\varphi\cos
\theta\mathbf{e}_{y}-\sin\theta\mathbf{e}_{z},\\
\mathbf{e}_{Z}  &  =\boldsymbol{\bar{\omega}}/|\boldsymbol{\bar{\omega}}%
|=\sin\theta\cos\varphi\mathbf{e}_{x}+\sin\theta\sin\varphi\mathbf{e}_{y}%
+\cos\theta\mathbf{e}_{z},
\end{align}
where $\theta$ $(\varphi$) is the polar (azimuth) anlge of $\boldsymbol{\bar
{\omega}}$ in the conventional coordinate $(\mathbf{e}_{x},\mathbf{e}%
_{y},\mathbf{e}_{z})$ with $\mathbf{e}_{z}$ along the N-V symmetry axis. The
nuclear spin dissipation include pure dephasing [the second term of Eq.
(\ref{ME_N})] due to the fluctuation of $\hat{F}_{Z}$ and relaxation [the last
two terms of Eq. (\ref{ME_N}), with $\hat{I}_{\pm}\equiv\hat{I}_{X}\pm
i\hat{I}_{Y}$] due to the fluctuation of $\hat{F}_{\pm}\equiv\hat{F}_{X}\pm
i\hat{F}_{Y}$. Typically the nuclear spin level splitting $|\boldsymbol{\bar
{\omega}}|\ll\gamma_{1},\gamma_{\varphi}$, so we obtain
\end{subequations}
\begin{subequations}
\label{GAMMA}%
\begin{align}
\Gamma_{\varphi}  &  =\frac{\tau_{e}^{2}}{2T}|(\boldsymbol{\omega}%
_{e}-\boldsymbol{\omega}_{g})_{Z}|^{2},\label{GAMMA_PHI}\\
\Gamma_{+}  &  =\Gamma_{-}=\frac{\tau_{e}^{2}}{4T}|(\boldsymbol{\omega}%
_{e}-\boldsymbol{\omega}_{g})_{\perp}|^{2}, \label{GAMMA_PM}%
\end{align}
where $\mathbf{O}_{\perp}\equiv O_{X}\mathbf{e}_{X}+O_{Y}\mathbf{e}_{Y}$ is
the component perpendicular to the nuclear spin quantization axis,
$T=1/R+1/(\gamma_{1}+R)$ is the duration of one electron hopping cycle
(excitation time $1/R$ and de-excitation time $1/(\gamma_{1}+R)$), and
\end{subequations}
\begin{equation}
\tau_{e}=\sqrt{\frac{R+\frac{\gamma_{1}\gamma_{\varphi}}{\gamma_{1}%
+\gamma_{\varphi}}+\pi\gamma_{1}^{2}\delta^{((\gamma_{1}+\gamma_{\varphi}%
)/2)}(\Delta)}{R+\gamma_{1}}}\frac{\sqrt{2}}{2R+\gamma_{1}}\approx\frac
{\sqrt{2}}{2R+\gamma_{1}} \label{SIGMAE}%
\end{equation}
is the uncertainty of the electron dwell time in the excited state in each
hopping cycle. Here the last step of Eq. (\ref{SIGMAE}) holds at room
temperature, where $\gamma_{\varphi}\sim10^{7}$ MHz is much larger than
typical $\gamma_{1},R$, and $\Delta$. Equation (\ref{GAMMA}) shows that
nuclear spin dissipation vanishes when $\boldsymbol{\omega}_{g}%
=\boldsymbol{\omega}_{e}$, simply because in this case the nuclear spin
precession frequency is not randomized by the optically induced electron hopping.

\subsection{Physical picture}

Equations (\ref{ME_N})-(\ref{SIGMAE}) not only provide an quantitative and
analytical description for the dissipative nuclear spin dynamics due to an
optically pumped NV center, but also have a physically transparent
interpretation that substantiates the previous analytical (but qualitative)
and numerical results \cite{JiangPRL2008}. For example, the pure dephasing
rate in Eq.\ (\ref{GAMMA_PHI}) is directly connected to the nuclear spin phase
diffusion process by the optically induced random hopping of the electron
between the ground state $|g\rangle$ and the excited state $|e\rangle$
\cite{JiangPRL2008}. To clearly see this, lets consider the phase accumulation
of the nuclear spin during an interval $[0,t]$. Suppose that during this
interval, the electron undergoes $N$ hopping cycles, and that during the $k$th
cycle, the electron stays in $|g\rangle$ for an interval $\tau_{k}$, so the
total dwell times in $|g\rangle$ and $|e\rangle$ are $\tau=\sum_{k=1}^{N}%
\tau_{k}$ and $t-\tau$, respectively, and the nuclear spin accumulates a phase
factor $e^{-i(\mathbf{a}_{g}+\gamma_{N}\mathbf{B)}_{Z}\tau-i(\mathbf{a}%
_{e}+\gamma_{N}\mathbf{B)}_{Z}(t-\tau)}$. For $t\gg T$, the number of hopping
cycle $N\approx t/T\gg1$, i.e., $\tau$ is the sum of many random variables
$\{\tau_{k}\}$, so $\tau$ obeys Gaussian distribution centered at $P_{g}t$
with a standard deviation $\sqrt{N}\tau_{e}$, where $\tau_{e}$ is the rms
fluctuation of each $\tau_{k}$. Averaging the phase factor over this Gaussian
distribution gives $e^{-i|\boldsymbol{\bar{\omega}}|t}e^{-\Gamma_{\varphi}t},$
where $\Gamma_{\varphi}$ coincides with Eq. (\ref{GAMMA_PHI}) as long as
$\tau_{e}$ is given in Eq. (\ref{SIGMAE}), e.g., at room temperature, for weak
pumping $R\ll\gamma_{1}$, the uncertainty $\tau_{e}\approx\sqrt{2}/\gamma_{1}$
of the dwell time in $|e\rangle$ is dominated by the uncertainty in the
spontaneous emission; while for saturated pumping, $\tau_{e}\approx1/(\sqrt
{2}R)$ is strongly suppressed by the rapid optically induced transition
between $|e\rangle$ and $|g\rangle$. The relaxation rate $\Gamma_{\pm}$ in Eq.
(\ref{GAMMA_PM}) can be understood in a similar way.

analytical results Eqs. (\ref{GAMMA}) provide a microscopic basis for the
previous model \cite{JiangPRL2008} and experimental observations
\cite{DuttScience2007,MaurerScience2012,LiNatCommun2013}, e.g., it clearly
shows the initial increase of the dissipation rates $\Gamma_{\varphi}%
,\Gamma_{\pm}\propto R$ under weak pumping $R\ll\gamma_{1}$ and the motional
narrowing $\Gamma_{\varphi},\Gamma_{\pm}\propto1/R$ under saturated pumping
$R\gg\gamma_{1}$. The former arises from the increase of $T$ with decreasing
$R$ under weak pumping, while the latter comes from both the decrease of
$\tau_{e}\sim1/R$ and $T\sim1/R$ under saturated pumping. Our analytical
formula also demonstrate the possibility \cite{JiangPRL2008} to control
$\Gamma_{\varphi}$ and $\Gamma_{\pm}$ by using the magnetic field to tune the
nuclear quantization axis $\mathbf{e}_{Z}\propto\boldsymbol{\bar{\omega}}$,
e.g., if we tune $\mathbf{e}_{Z}$ to be perpendicular (parallel) to
$\boldsymbol{\omega}_{g}-\boldsymbol{\omega}_{e}$, then we can eliminate
nuclear spin pure dephasing (relaxation). Interestingly, the sum rule
\begin{equation}
\Gamma_{\varphi}+\Gamma_{+}+\Gamma_{-}=\frac{\tau_{e}^{2}}{2T}%
|\boldsymbol{\omega}_{e}-\boldsymbol{\omega}_{g}|^{2}\label{SUM_RULE}%
\end{equation}
suggests that reducing $\Gamma_{\varphi}$ ($\Gamma_{\pm}$) inevitably
increases $\Gamma_{\pm}$ ($\Gamma_{\varphi}$) and it is impossible to suppress
$\Gamma_{\varphi}$ and $\Gamma_{\pm}$ simultaneously, unless the NV states are
tuned such that $\boldsymbol{\omega}_{g}=\boldsymbol{\omega}_{e}$.

\subsection{Connection to experimental observations}

Equation (\ref{ME_N}) describes the dissipative evolution of the nuclear spin
in the tilted coordinate $(\mathbf{e}_{X},\mathbf{e}_{Y},\mathbf{e}_{Z})$ with
$\mathbf{e}_{Z}\propto\boldsymbol{\bar{\omega}}$. From Eq. (\ref{ME_N}), we
obtain the Bloch equations
\begin{subequations}
\label{EOM_AVEI}%
\begin{align}
\partial_{t}\langle\hat{I}_{Z}\rangle &  =-\frac{\langle\hat{I}_{Z}\rangle
}{T_{1}},\\
\partial_{t}\langle\hat{I}_{+}\rangle &  =(i|\boldsymbol{\bar{\omega}}%
|-\frac{1}{T_{2}})\langle\hat{I}_{+}\rangle,
\end{align}
for the average nuclear spin $\langle\hat{\mathbf{I}}(t)\rangle\equiv
\operatorname*{Tr}\hat{\mathbf{I}}\hat{p}(t)$, where $T_{1}=1/(\Gamma
_{+}+\Gamma_{-})$ and $T_{2}=1/(\Gamma_{\varphi}+(\Gamma_{+}+\Gamma_{-})/2)$.
Then the sum rule in Eq. (\ref{SUM_RULE}) implies $1/T_{2}+1/(2T_{1}%
)\propto|\boldsymbol{\omega}_{e}-\boldsymbol{\omega}_{g}|^{2}$, i.e., tuning
the magnetic field can prolong $T_{1}$ time ($T_{2}$ time) at the cost of
reducing $T_{2}$ time ($T_{1}$ time).

The above Bloch equations have simple solutions $\langle\hat{I}_{Z}%
(t)\rangle=\langle\hat{I}_{Z}(0)\rangle e^{-t/T_{1}}$ and $\langle\hat{I}%
_{+}(t)\rangle=e^{i|\boldsymbol{\bar{\omega}}|t}e^{-t/T_{2}}\langle\hat{I}%
_{+}(0)\rangle$. However, nuclear spin initialization and measurement are
usually performed in the conventional coordinate $(\mathbf{e}_{x}%
,\mathbf{e}_{y},\mathbf{e}_{z})$ with $\mathbf{e}_{z}$ along the N-V axis, so
$T_{1}$ and $T_{2}$ will be mixed in the observed signals. For example, Dutt
\textit{et al.}\cite{DuttScience2007} initialize a strongly coupled $^{13}$C
nuclear spin-1/2 (hereafter referred to as $^{13}\mathrm{C}_{\mathrm{b}}$,
according to the notation of Gali \cite{GaliPRB2009a}) into the eigenstate
$(|\uparrow\rangle+|\downarrow\rangle)/\sqrt{2}$ of $\hat{I}_{x}$, let it
evolve freely for an interval $\tau$, and then measure $\hat{I}_{x}$ through a
$\pi/2$ pulse $e^{-i\pi\hat{I}_{y}/2}$ followed by a fluorescence readout of
$\hat{I}_{z}$ via the NV center. According to Eq. (\ref{XYZ}), the measured
signal $\langle\hat{I}_{x}(t)\rangle=\sum_{\alpha=X,Y,Z}(\mathbf{e}_{x}%
\cdot\mathbf{e}_{\alpha})\langle\hat{I}_{\alpha}(t)\rangle$ consists of a
non-oscillatory term $e^{-t/T_{1}}\sin^{2}\theta\cos^{2}\varphi/2$ that decays
with a time scale $T_{1}$ and an oscillatory term $e^{-t/T_{2}}(1-\cos
^{2}\varphi\sin^{2}\theta)\cos(|\boldsymbol{\bar{\omega}}|t)/2$ that decays
with a time scale $T_{2}$. The oscillating feature has been observed experimentally
\cite{DuttScience2007}. When the nuclear spin quantization axis $\mathbf{e}%
_{Z}$ is parallel to the initial state polarization direction $\mathbf{e}_{x}%
$, i.e., $\theta=\pi/2$ and $\varphi=0$, the oscillatory feature disappears.

At room temperature, when the magnetic field is along the $z$ axis, the
optical transition is spin conserving. The cyclic transition between the $m=0$
ground state $|g\rangle=|0_{g}\rangle$ and excited state $|e\rangle
\equiv|0_{e}\rangle$ does not contribute to nuclear spin dissipation since
$\boldsymbol{\omega}_{g}=\boldsymbol{\omega}_{e}=0$. When $\mathbf{B}$
deviates from the $z$ axis, its transverse component $\mathbf{B}_{\mathrm{T}%
}\equiv B_{x}\mathbf{e}_{x}+B_{y}\mathbf{e}_{y}$ mixes the $m=0$ sublevels and
the $m=\pm1$ sublevels, so that $\boldsymbol{\omega}_{g}=-(2\gamma
_{e}/D_{\mathrm{gs}})\mathbf{B}_{\mathrm{T}}\cdot\mathbf{A}_{g}$ and
$\boldsymbol{\omega}_{e}=-(2\gamma_{e}/D_{\mathrm{es}})\mathbf{B}_{\mathrm{T}%
}\cdot\mathbf{A}_{e}$, where $D_{\mathrm{gs}}$ ($D_{\mathrm{es}}$) is the
zero-field splitting in the NV ground (excited) state and $\gamma
_{e}=28.025\ $\textrm{MHz/mT} is the gyromagnetic ratio of the NV electron.
This can be understood as a hyperfine enhancement of the nuclear spin g-factor
\cite{DuttScience2007} (see the next section for more detailed discussion). As
a result, the nuclear spin dissipation rates $\Gamma_{\varphi},\Gamma_{\pm}$
are proportional to $|\mathbf{B}_{\mathrm{T}}|^{2}$, as observed
experimentally \cite{DuttScience2007,JiangPRL2008}. For the $^{13}%
$C$_{\mathrm{b}}$ nucleus studied by Dutt \textit{et al}.
\cite{DuttScience2007}, the HFI tensor has been obtained by ab initio
calculations \cite{GaliPRB2009a} as
\end{subequations}
\begin{subequations}
\label{HFI}%
\begin{align}
\mathbf{A}_{g}(^{13}\mathrm{C}_{b}) &  \approx%
\begin{bmatrix}
-8 & 0 & -0.7\\
0 & -8.99 & 0\\
-0.7 & 0 & -8.00
\end{bmatrix}
\ \mathrm{MHz},\\
\mathbf{A}_{e}(^{13}\mathrm{C}_{b}) &  \approx%
\begin{bmatrix}
-3.78 & 0.19 & -1.47\\
0.19 & -5.83 & 0.22\\
-1.47 & 0.22 & -4.12
\end{bmatrix}
\ \mathrm{MHz},
\end{align}
From this HFI tensor, we estimate $|\boldsymbol{\omega}_{g}-\boldsymbol{\omega
}_{e}|\sim0.3$ $\mathrm{MHz}$ when $|\boldsymbol{\omega}_{g}|=1\ \mathrm{MHz}%
$. Under optical pumping rate $R=\gamma_{1}$ (the experimental condition
\cite{DuttScience2007}), the two-level fluctuator model [Eq. (\ref{GAMMA_PHI}%
)] gives a nuclear spin dephasing rate $1/T_{2}\approx\Gamma_{\varphi}%
\approx(250\ \mathrm{\mu s})^{-1}$ [$\ll1/\tau_{\mathrm{NV}}\sim
(4\ \mathrm{ns})^{-1}$, so the NV center is a good Markovian bath], which is
two orders of magnitudes smaller than the experimentally observed value
$\sim(1\ \mathrm{\mu s})^{-1}$. Equivalently, to be consistent with the
experiment \cite{DuttScience2007}, the difference $|\boldsymbol{\omega}%
_{g}-\boldsymbol{\omega}_{e}|$ must be assumed to be 10 times larger
\cite{JiangPRL2008}. This large discrepancy suggests that the leakage from
$m=0$ subspace to $m=\pm1$ subspace may plays an important role in determining
the nuclear spin dissipation. In the next section, we shall demonstrate that
the small leakage from $|0\rangle$ subspace to the $|\pm1\rangle$ subspace
could introduce additional contributions that may dominates the nuclear spin
dissipation.

\section{Seven-level fluctuator model: numerical and analytical results}

\subsection{Model}

\begin{figure}[ptb]
\includegraphics[width=0.8\columnwidth]{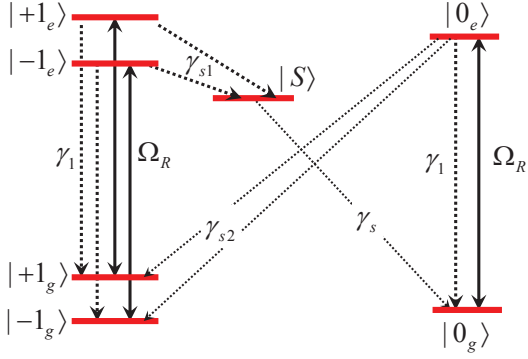}\caption{Sketch of the
seven energy levels of the NV center under optical pumping.}%
\label{energylevel}
\end{figure}

Now we consider general optical pumping of the NV center at \textit{room
temperature}, incorporating the finite intersystem crossing between $m=0$ and
$m=\pm1$ subspaces. In this case, the electron-nuclear coupled system still
obeys Eq. (\ref{ME0}). The only difference is that now the Liouville
superoperator $\mathcal{L}_{e}$ for the optically pumped NV center includes
seven energy levels: the ground triplet $|m\rangle|g\rangle\equiv|m_{g}%
\rangle$ (i.e., $|0_{g}\rangle,|\pm1_{g}\rangle$), the excited triplet
$|m\rangle|e\rangle\equiv|m_{e}\rangle$ (i.e., $|0_{e}\rangle,|\pm1_{e}%
\rangle$), and the metastable singlet $|S\rangle$, where $|g\rangle$
($|e\rangle$) denote the ground (excited) orbital. The unitary part of
$\mathcal{L}_{e}$ is a seven-level NV Hamiltonian (However, we haven't
consider the effect of ionization of the NV center which would happen when the
laser intensity is very strong \cite{MaurerScience2012})
\end{subequations}
\[
\hat{H}_{e}=\Delta\hat{\sigma}_{e,e}+\hat{H}_{\mathrm{gs}}+\hat{H}%
_{\mathrm{es}}+\frac{\Omega_{R}}{2}(\hat{\sigma}_{e,g}+h.c.),
\]
where $\Delta$ is the optical detuning between the zero-phonon line and the
laser frequency, $\hat{H}_{\mathrm{gs}}=D_{\mathrm{gs}}\hat{S}_{g,z}%
^{2}+\gamma_{e}\mathbf{B}\cdot\hat{\mathbf{S}}_{g}$ and $\hat{H}_{\mathrm{es}%
}=D_{\mathrm{es}}\hat{S}_{e,z}^{2}+\gamma_{e}\mathbf{B}\cdot\hat{\mathbf{S}%
}_{e}$ describe, respectively, the ground state triplet with zero-field
splitting $D_{\mathrm{gs}}=2.87$ $\mathrm{GHz}$ and the excited state triplet
with zero-field splitting $D_{\mathrm{es}}=1.41\ \mathrm{GHz}$. The
dissipative part of $\mathcal{L}_{e}$ includes various dissipation processes
between the seven levels of the NV center as sketched in Fig.
\ref{energylevel}: the spontaneous emission from the excited orbital
$|e\rangle$ to the ground orbital $|g\rangle$ with rate $\gamma_{1}%
=1/(12\ \mathrm{ns})$ \cite{FuchsPRL2012}, the non-radiative decay from
$|\pm1_{e}\rangle$ to the metastable singlet $|S\rangle$ with rate
$\gamma_{s1}\approx\gamma_{1}$ followed by the non-radiative decay from
$|S\rangle$ to $|0_{g}\rangle$ with rate $\gamma_{s}=1/(143\ \mathrm{ns}%
)$\cite{AcostaPRB2010}, the leakage from $|0_{e}\rangle$ to $|\pm1_{g}\rangle$
with equal rates $\gamma_{s2}\ll\gamma_{s1}$, and the orbital dephasing of the
excited state $\hat{L}=\hat{\sigma}_{e,e}$ with rate $\gamma_{\varphi}%
\sim10^{7}\ \mathrm{MHz}$ \cite{AbtewPRL2011,FuPRL2009}.

As discussed in the previous section, there are two processes contributing to
nuclear spin dissipation. The one involving the electron spin flip is strongly
suppressed away from the NV center ground state and excited state level
anticrossing. Thus, in our analytical derivation below, we consider the other
process that does not change the electron spin state, i.e., we drop the
off-diagonal electron spin flip terms in $\hat{\mathbf{F}}$ and only keep the
diagonal part. The magnetic field component $\mathbf{B}_{\mathrm{T}}\equiv
B_{x}\mathbf{e}_{x}+B_{y}\mathbf{e}_{y}$ perpendicular to the N-V axis ($z$
axis) slightly shifts the electron levels and mixes the electron states from
$|m_{g}\rangle$ and $|m_{e}\rangle$ to $|\tilde{m}_{g}\rangle$ and $|\tilde
{m}_{e}\rangle$ $(m=0,\pm1)$. For $\gamma_{e}|\mathbf{B}_{\mathrm{T}}|\ll
D_{\mathrm{gs}},D_{\mathrm{es}}$, the level shift can be safely neglected, but
the state mixing has a nontrivial influence on the diagonal part of
$\hat{\mathbf{F}}$, i.e., we need to keep the terms diagonal in the mixed
basis $|\tilde{m}_{g}\rangle$ and $|\tilde{m}_{e}\rangle$:
\[
\hat{\mathbf{F}}\approx\sum_{m}(\hat{\sigma}_{\tilde{m}_{g},\tilde{m}_{g}%
}\langle\tilde{m}_{g}|\hat{\mathbf{S}}_{g}|\tilde{m}_{g}\rangle\cdot
\mathbf{A}_{g}+\hat{\sigma}_{\tilde{m}_{e},\tilde{m}_{e}}\langle\tilde{m}%
_{e}|\hat{\mathbf{S}}_{e}|\tilde{m}_{e}\rangle\cdot\mathbf{A}_{e}).
\]
Up to the first order of the small quantities $|\gamma_{e}\mathbf{B}%
_{\mathrm{T}}|/D_{\mathrm{gs}}$ and $|\gamma_{e}\mathbf{B}_{\mathrm{T}%
}|/D_{\mathrm{es}}$, we obtain (hereafter $|m_{g/e}\rangle$ stands for
$|\tilde{m}_{g/e}\rangle$):
\begin{align*}
\langle0_{g}|\hat{\mathbf{S}}_{g}|0_{g}\rangle &  \approx-\frac{2\gamma_{e}%
}{D_{\mathrm{gs}}}\mathbf{B}_{\mathrm{T}},\\
\langle0_{e}|\hat{\mathbf{S}}_{e}|0_{e}\rangle &  \approx-\frac{2\gamma_{e}%
}{D_{\mathrm{es}}}\mathbf{B}_{\mathrm{T}},\\
\langle\pm1_{g}|\hat{\mathbf{S}}_{g}|\pm1_{g}\rangle &  \approx\pm
\mathbf{e}_{z}+\frac{\gamma_{e}}{D_{\mathrm{gs}}}\mathbf{B}_{\mathrm{T}},\\
\langle\pm1_{e}|\hat{\mathbf{S}}_{e}|\pm1_{e}\rangle &  \approx\pm
\mathbf{e}_{z}+\frac{\gamma_{e}}{D_{\mathrm{es}}}\mathbf{B}_{\mathrm{T}}.
\end{align*}
The terms proportional to $\mathbf{B}_{\mathrm{T}}$ lead to hyperfine
enhancement of the nuclear spin g-factor \cite{DuttScience2007}, e.g., the
term $\hat{\sigma}_{0_{g},0_{g}}\langle0_{g}|\hat{\mathbf{S}}_{g}|0_{g}%
\rangle\cdot\mathbf{A}_{g}$ in $\hat{\mathbf{F}}$ can be written as
$\hat{\sigma}_{0_{g},0_{g}}\gamma_{N}\mathbf{B}_{\mathrm{T}}\cdot
\lbrack-2\gamma_{e}\mathbf{A}_{g}/(\gamma_{N}D_{\mathrm{gs}})]$, where
$[\cdots]$ is the correction to the nuclear gyromagnetic ratio by the HFI
conditioned on the electron being in $|0_{g}\rangle$. Since $\gamma
_{e}|\mathbf{B}_{\mathrm{T}}|\ll D_{\mathrm{gs}},D_{\mathrm{es}}$, the
hyperfine enhancement terms in $\langle\pm1_{g}|\hat{\mathbf{S}}_{g}|\pm
1_{g}\rangle$ and $\langle\pm1_{e}|\hat{\mathbf{S}}_{e}|\pm1_{e}\rangle$ can
be safely dropped, so that%
\[
\hat{\mathbf{F}}\approx\hat{S}_{g,z}\mathbf{b}_{g}+\hat{S}_{e,z}\mathbf{b}%
_{e}+\hat{\sigma}_{0_{g},0_{g}}\mathbf{a}_{g}+\hat{\sigma}_{0_{e},0_{e}%
}\mathbf{a}_{e},
\]
where $\mathbf{a}_{g}=-(2\gamma
_{e}/D_{\mathrm{gs}})\mathbf{B}_{\mathrm{T}}\cdot\mathbf{A}_{g}$,
$\mathbf{a}_{e}=-(2\gamma_{e}/D_{\mathrm{es}})\mathbf{B}_{\mathrm{T}}\cdot\mathbf{A}_{e}$
and $\mathbf{b}_{g/e}=\mathbf{e}_{z}\cdot\mathbf{A}_{g/e}$.

Now the second term of Eq. (\ref{ME0}) describes the nuclear spin precession
conditioned on the electron state: the precession frequency is $\gamma
_{N}{\mathbf{B}}\pm\mathbf{b}_{g}$ ($\gamma_{N}{\mathbf{B}}\pm\mathbf{b}_{e}$)
when the electron state is $|\pm1_{g}\rangle$ ($|\pm1_{e}\rangle$), or
$\gamma_{N}{\mathbf{B}}\pm\mathbf{a}_{g}$ ($\gamma_{N}{\mathbf{B}}%
\pm\mathbf{a}_{e}$) when the electron state is $|0_{g}\rangle$ ($|0_{e}%
\rangle$), or $\gamma_{N}{\mathbf{B}}$ when the electron is in the metastable
singlet $|S\rangle$. The optically induced hopping of the electron between
different states randomizes the precession frequency and leads to nuclear spin
dissipation \cite{JiangPRL2008}. Below we derive analytically a closed
equation of motion of the nuclear spin to describe these effects.

\subsection{Lindblad master equation for nuclear spin}

The time scale $\tau_{\mathrm{NV}}=1/\min\{\gamma_{s2,}R\}$ for the
seven-level NV center to reach its steady state is determined by the time
scale of the slowest process: the intersystem crossing from $m=0$ subspace to
$m=\pm1$ subspaces if the optical pumping is strong, or the optical pumping
rate $R=\Omega_{R}^{2}/\gamma_{\varphi}$ from the ground orbital to the
excited orbital if the optical pumping is weak. When $\tau_{\mathrm{NV}}$ is
much shorter than the time scale $T_{1},T_{2}$ of the nuclear spin
dissipation, we can follow exactly the same procedures as used in the previous
section to derive the Lindblad master equation for the nuclear spin density
matrix and the Bloch equation for the average nuclear spin angular momentum. The former
(latter) has exactly the same form as Eq. (\ref{ME_N}) [Eq. (\ref{EOM_AVEI})]
and describes the nuclear spin dissipation in the tilted cartesian coordinate
$(\mathbf{e}_{X},\mathbf{e}_{Y},\mathbf{e}_{Z})$ with $\mathbf{e}_{Z}%
\equiv\boldsymbol{\bar{\omega}}/|\boldsymbol{\bar{\omega}}|$ [see Eq.
(\ref{XYZ})] and $\boldsymbol{\bar{\omega}}=\gamma_{N}\mathbf{B}+P_{0_{g}%
}\mathbf{a}_{g}+P_{0_{e}}\mathbf{a}_{e}$, where $P_{0_{g}}$ and $P_{0_{e}}$
are steady state populations of the NV center on $|0_{g}\rangle$ and
$|0_{e}\rangle$. The detail expression of steady populations is given in
appendix A.

Now we discuss the analytical expressions for the nuclear spin pure dephasing
rate $\Gamma_{\varphi}$ and relaxation rate $\Gamma_{\pm}$. The former comes
from the fluctuation of $\hat{F}_{Z}$, while the latter comes from the
fluctuation of $\hat{F}_{\pm}\equiv\hat{F}_{X}\pm i\hat{F}_{Y}$. Since
$\hat{\mathbf{F}}$ is a linear combination of $\hat{S}_{g,z},\hat{S}%
_{e,z},\hat{\sigma}_{0_{g},0_{g}},$ and $\hat{\sigma}_{0_{e},0_{e}}$, the
fluctuation of $\hat{F}_{Z}$ and $\hat{F}_{\pm}$ involve various
cross-correlations among these four operators. Fortunately, due to the large
orbital dephasing at room temperature, the optical pumping rate from the
ground orbital to the excited orbital is nearly independent of the spin state and 
the coherence between electron states can be neglected.
This allows us to neglect the cross correlation between $\{\hat{S}_{g,z}%
,\hat{S}_{e,z}\}$ and $\{\hat{\sigma}_{0_{g},0_{g}},\hat{\sigma}_{0_{e},0_{e}%
}\}$ (see Appendix B for details). So $\Gamma_{\varphi}$ and $\Gamma_{\pm}$
are the sum of the contributions $\Gamma_{\varphi}^{(1)},\Gamma_{\pm}^{(1)}$
from the fluctuation of $\hat{S}_{g,z},\hat{S}_{e,z}$ associated with the
$m=\pm1$ subspace and the contributions $\Gamma_{\varphi}^{(0)}$, $\Gamma
_{\pm}^{(0)}$ from the fluctuation of $\hat{\sigma}_{0_{g},0_{g}},\hat{\sigma
}_{0_{e},0_{e}}$ associated with the $m=0$ subspace. Unless explicitly
specified, hereafter we consider a typical situation $\left\vert
\boldsymbol{\bar{\omega}}\right\vert \ll R$, $\gamma_{s1}$.

The contribution from $m=\pm1$ subspace is%
\begin{align}
\Gamma_{\varphi}^{(1)}  &  =\frac{2P_{-1_{e}}}{\gamma_{s1}}\left[  \left(
b_{e,Z}+b_{g,Z}\frac{\gamma_{1}+\gamma_{s1}+R}{R}\right)  ^{2}-b_{g,Z}%
b_{e,Z}\frac{\gamma_{s1}}{R}\right]  ,\label{pure_pm1}\\
\Gamma_{\pm}^{(1)}  &  \approx\frac{P_{-1_{e}}}{\gamma_{s1}}\left[  \left(
\mathbf{b}_{e,\perp}+\frac{\gamma_{1}+\gamma_{s1}+R}{R}\mathbf{b}_{g,\perp
}\right)  ^{2}-(\mathbf{b}_{g,\perp}\cdot\mathbf{b}_{e,\perp})\frac
{\gamma_{s1}}{R}\right]  , \label{relax_pm1}%
\end{align}
where $P_{i}$ is the steady-state population of the electron state $|i\rangle
$. Formally $\Gamma_{\varphi}^{(1)}$ and $\Gamma_{\pm}^{(1)}$ are independent
of the magnetic field, but actually the components $b_{g,Z},\mathbf{b}%
_{g,\perp}\equiv b_{g,X}\mathbf{e}_{X}+b_{g,Y}\mathbf{e}_{Y}$, etc. are
defined in the tilted coordinate $\mathbf{e}_{X},\mathbf{e}_{Y},\mathbf{e}%
_{Z}$ [see Eqs. (\ref{XYZ})], which in turn depends on the magnetic field.
Importantly, $\Gamma_{\varphi}^{(1)}$ and $\Gamma_{\pm}^{(1)}$ do not vanish
even in zero magnetic field. This provides a possible solution to the puzzling
observation of nuclear spin dephasing in zero magnetic field
\cite{DuttScience2007}, which has been speculated to be due to the orbital
fluctuation of the NV center in the excited state \cite{JiangPRL2008}.

Equations (\ref{pure_pm1}) and (\ref{relax_pm1}) exhibit four features. First,
$\Gamma_{\varphi}^{(1)}$ and $\Gamma_{\pm}^{(1)}$ do not vanish when
$\mathbf{b}_{g}=\mathbf{b}_{e}$ and $\mathbf{a}_{g}=\mathbf{a}_{e}$, as
opposed to the two-level fluctuator model [Eq. (\ref{GAMMA})]. This is because
in the two-level fluctuator model, the electron only hops between $|g\rangle$
(with nuclear spin precession frequency $\gamma_{N}\mathbf{B}+\mathbf{a}_{g}%
$)\ and $|e\rangle$ (with nuclear spin precession frequency $\gamma
_{N}\mathbf{B}+\mathbf{a}_{e}$):\ when $\mathbf{a}_{g}=\mathbf{a}_{e}$, the
electron hopping does not randomize the nuclear spin precession, so there is
no nuclear spin dissipation. By contrast, in the seven-level fluctuator model,
the electron can hop between seven energy levels, each of which corresponds to
a different nuclear spin precession frequency (see the discussion at the end
of the previous subsection). Therefore, even if the hyperfine of the
excited state is the same as that of the ground state, the dissipation process
also exists. This conclusion is different from the expectation
\cite{PfaffNatPhys2013,BlokNatPhys2014} that the decoherence comes from the
hyperfine difference between the ground state and excited state. The nuclear
spin dissipation vanishes only when all these precession frequencies are
equal, i.e., when $\mathbf{a}_{g}=\mathbf{a}_{e}=\mathbf{b}_{g}=\mathbf{b}%
_{e}=0$.  Second,$\Gamma_{\pm}^{(1)}$ and $\Gamma_{\varphi}^{(1)}$ are
proportional to the electron population $P_{-1_{e}}=P_{+1_{e}}\propto
\gamma_{s2}$ in the $|\pm1_{e}\rangle$ level, which vanishes when the leakage
rate $\gamma_{s2}$ from $m=0$ subspace to $m=\pm1$ subspace vanishes. Third,
under weak pumping $R\ll\gamma_{1},\gamma_{s1}$, we have $\Gamma_{\varphi
}^{(1)},\Gamma_{\pm}^{(1)}\propto1/R$ increasing with decreasing pumping
strength, until the pumping is too weak for the Markovian assumption
$\tau_{\mathrm{NV}}\ll T_{1},T_{2}$, based on which our analytical formula are
derived, to remain valid. Upon further decrease of the pumping strength, the
NV center becomes a non-Markovian bath and the nuclear spin dissipation rates
would show a maximum and then decrease (see the next subsection for more
discussions). Finally, under saturated optical pumping, $\Gamma_{\varphi
}^{(1)}$ and $\Gamma_{\pm}^{(1)}$ are \textit{saturated} instead of being
suppressed:
\begin{subequations}
\label{pm1_sat}%
\begin{align}
\Gamma_{\varphi}^{(1)} &  \approx2\times\frac{\tilde{\tau}_{1}^{2}}{2\tilde
{T}}\left(  \frac{b_{g,Z}+b_{e,Z}}{2}\right)  ^{2},\label{pure_pm1_sat}\\
\Gamma_{\pm}^{(1)} &  \approx2\times\frac{\tilde{\tau}_{1}^{2}}{4\tilde{T}%
}\left\vert \left(  \frac{\mathbf{b}_{g}+\mathbf{b}_{e}}{2}\right)  _{\perp
}\right\vert ^{2},\label{relax_pm1_sat}%
\end{align}
where $\tilde{T}\equiv2/\gamma_{s1}+1/\gamma_{s2}+1/\gamma_{s}\approx
1/\gamma_{s2}$ is the average duration of one electron hopping cycle,
$\tilde{\tau}_{1}\equiv2/\gamma_{s1}$ is the uncertainty of the dwell
time in the $|+1\rangle$($|-1\rangle$) level, and the prefactor $2$ accounts
for the contribution from the $m=+1$ and $m=-1$ subspaces.

Equations (\ref{pm1_sat}) can be understood as follows. First, under strong
pumping, the hopping time between the ground orbital and the excited orbital
is negligibly small, so the $m=+1$ (or $m=-1$) subspace effectively becomes a
single energy level with nuclear spin precession frequency $\gamma
_{N}\mathbf{B}+(\mathbf{b}_{g}+\mathbf{b}_{e})/2$(or $\gamma_{N}%
\mathbf{B}-(\mathbf{b}_{g}+\mathbf{b}_{e})/2$). Second, the duration
$\tilde{T}\approx1/\gamma_{s2}$ of one hopping cycle is ultimately limited by
the slowest process:\ the intersystem crossing from $m=0$ to $m=\pm1$
subspace. Therefore, Eqs. (\ref{pm1_sat}) correspond to an effective two-level
fluctuator model [cf. Eqs. (\ref{GAMMA})]: one state is the $m=+1$ (or
$m=-1$) subspace with nuclear spin precession frequency $\gamma_{N}%
\mathbf{B}+(\mathbf{b}_{g}+\mathbf{b}_{e})/2$ (or $\gamma_{N}\mathbf{B}%
-(\mathbf{b}_{g}+\mathbf{b}_{e})/2$) and the other state is the subspace excluding
$\vert\pm1\rangle$ subspace, which produce a nuclear spin precession frequency $\gamma
_{N}\mathbf{B}$. Although strong optical pumping suppresses the randomization
of the nuclear spin precession due to spin-conserving electron hopping between
the ground orbital and excited orbital inside the $m=+1$ (or $m=-1$) subspace,
there is extra contribution due to the random electron hopping between the
$m=\pm1$ subspace and the subspace excluding $\vert \pm1\rangle$ subspace.


The contributions from $m=0$ subspace involve $\mathbf{a}_{g}$ and
$\mathbf{a}_{e}$ in a quadratic form, so $\Gamma_{\varphi}^{(0)},\Gamma_{\pm
}^{(0)}\propto|\mathbf{B}_{\mathrm{T}}|^{2}$ increases significantly with the
magnetic field components perpendicular to the N-V axis. Due to the finite
leakage from $m=0$ into $m=\pm1$ subspace, the analytical expressions for
$\Gamma_{\varphi}^{(0)}$ and $\Gamma_{\pm}^{(0)}$ are very tedious (see
Appendix B), so here we discuss the limits of weak pumping and strong pumping.
Under weak pumping, $\Gamma_{\varphi}^{(0)},\Gamma_{\pm}^{(0)}\propto1/R$
decrease with increasing pumping strength (this behavior does not persists
down to $R\ll1/T_{1}$ or $1/T_{2}$, where our Markovian assumption does not
hold). Under strong optical pumping, they are saturated:
\end{subequations}
\begin{subequations}
\label{GAMMA_SAT}%
\begin{align}
\Gamma_{\varphi}^{(0)}  &  =\frac{\tilde{\tau}_{0}^{2}}{2\tilde{T}}\left\vert
\left(  \frac{\mathbf{a}_{g}+\mathbf{a}_{e}}{2}\right)  _{Z}\right\vert
^{2},\\
\Gamma_{\pm}^{(0)}  &  =\frac{\tilde{\tau}_{0}^{2}}{4\tilde{T}}\left\vert
\left(  \frac{\mathbf{a}_{g}+\mathbf{a}_{e}}{2}\right)  _{\perp}\right\vert
^{2},
\end{align}
where
\end{subequations}
\[
\tilde{\tau}_{0}=\frac{\sqrt{2}}{\gamma_{s2}\tilde{T}}\sqrt{\frac{2}%
{\gamma_{s1}^{2}}+\frac{1}{\gamma_{s1}\gamma_{s}}+\frac{1}{2\gamma_{s}^{2}}%
}\overset{\gamma_{s}\ll\gamma_{s1}}{\approx}\frac{1}{\gamma_{s}}%
\]
is the uncertainty of the time for the electron dwelling at the $m=0$
subspace. Similar to the contributions from the $m=\pm1$ subspace, under
strong optical pumping, the contributions from the $m=0$ subspace correspond
to an effective two-level fluctuator model:
one is the $m=0$ subspace with nuclear spin precession frequency
$\gamma_{N}\mathbf{B+(a}_{g}+\mathbf{a}_{e})/2$, the other state is the subspace
excluding the $m=0$ subspace, which produce a nuclear spin precession frequency
$\gamma_{N}\mathbf{B}$.

When the leakage from $m=0$ subspace to $m=\pm1$ subspace is neglected (i.e.,
$\gamma_{s2}=0$), the steady-state populations in the $m=\pm1$ subspace
vanish, corresponding to perfect optical initialization of the NV center into
the state $|0_{g}\rangle$. In this case, we have $\Gamma_{\varphi}%
^{(1)}=\Gamma_{\pm}^{(1)}=0$ and
\begin{subequations}
\label{GAMMA_TLS}%
\begin{align}
\Gamma_{\varphi}^{(0)}  &  =\frac{P_{0_{e}}P_{0_{g}}}{2R+\gamma_{1}}%
(a_{g,Z}-a_{e,Z})^{2},\\
\Gamma_{\pm}^{(0)}  &  \approx\frac{1}{2}\frac{P_{0_{e}}P_{0_{g}}}%
{2R+\gamma_{1}}(\mathbf{a}_{g,\perp}-\mathbf{a}_{e,\perp})^{2},
\end{align}
where $P_{0_{e}}=1-P_{0_{g}}=R/(2R+\gamma_{1})$. This recover the
room-temperature two-level fluctuator model [Eqs. (\ref{GAMMA}) and
(\ref{SIGMAE})]. This can be easily understood: since the population is
trapped in the $m=0$ subspace, the fluctuation of the nuclear spin precession
frequency could only come from the difference between $\mathbf{a}_{g}$ and
$\mathbf{a}_{e}$.

Below we discuss two situations: (i) the magnetic field is along the N-V axis
($z$ axis); (ii) the magnetic field is perpendicular to the N-V axis ($z$ axis).

\subsection{Magnetic field along N-V axis ($z$ axis)}

When the magnetic field is along the N-V symmetric axis ($z$ axis), we have
$\mathbf{a}_{g}=\mathbf{a}_{e}=0$, so the average precession frequency
$\boldsymbol{\bar{\omega}}=\gamma_{N}\mathbf{B}$ is along the $-z$ axis, and
the tilted axis $(\mathbf{e}_{X},\mathbf{e}_{Y},\mathbf{e}_{Z})$ can be
chosen as $(\mathbf{e}_{x},-\mathbf{e}_{y},-\mathbf{e}_{z})$.
Since $\mathbf{a}_{g}=\mathbf{a}_{e}=0$, only $m=\pm1$ subspace contribute to
nuclear spin dissipation: $\Gamma_{\varphi}=\Gamma_{\varphi}^{(1)}$ and
$\Gamma_{\pm}=\Gamma_{\pm}^{(1)}$ [see Eqs. (\ref{pure_pm1}) and
(\ref{relax_pm1})].

\begin{figure}[ptb]
\includegraphics[width=\columnwidth]{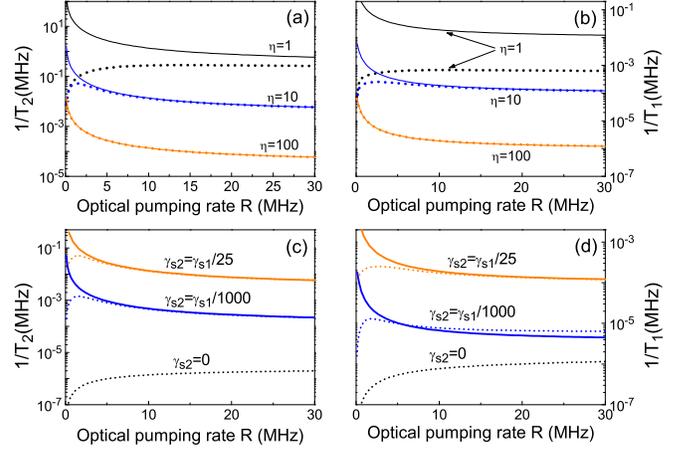}\caption{Comparison of
analytical (solid lines) and exact numerical results (dashed lines) for
nuclear spin $1/T_{2}$ [(a) and (c)] and $1/T_{1}$ [(b) and (d)] in a magnetic
field $B_{z}=5$ mT along the N-V axis as functions of the optical pumping rate
$R$. Relevant parameters are $\mathbf{A}_{g/e}=\mathbf{A}_{g/e}(^{13}%
\mathrm{C}_{b})/\eta$ ($\eta=1,$ 10, 100), and $\gamma_{s2}=\gamma_{s1}/25$ in
(a) and (b); $\mathbf{A}_{g/e}=\mathbf{A}_{g/e}(^{13}\mathrm{C}_{b})/10$ and
$\gamma_{s2}=0,\gamma_{s1}/1000,\gamma_{s1}/25$ in (c) and (d).}%
\label{ana_numer}
\end{figure}

To begin with, we demonstrate the validity of our analytical formula Eqs.
(\ref{pure_pm1}) and (\ref{relax_pm1}) by comparing them with the exact
numerical results from directly solving the electron-nuclear coupled equations
of motion [Eq. (\ref{ME0})]. we estimate the typical nuclear spin dissipation
time $\sim T_{1},T_{2}\ll\tau_{\mathrm{NV}}$ for $^{13}C_{b}$. In this case,
the NV center is a highly non-Markovian bath beyond the description of our
analytical formula. To see how our analytical fomula becomes progressively
applicable when going from the non-Markovian regime to the Markovian regime,
we manually scale down $\mathbf{A}_{g}$ and $\mathbf{A}_{e}$ by a factor
$\eta=1,10$, and $100$ to decrease the nuclear spin dissipation. Figures
\ref{ana_numer}(a) and \ref{ana_numer}(b) show three features: (i) Both the
exact results (dashed lines) and our analytical results (solid lines) tend to
saturate at large $R$, even for very strong HFI ($\eta=1$), where the NV
center is highly non-Markovian. (ii) With increasing $\eta$ and/or optical
pumping rate $R$, the nuclear spin dissipation rates $1/T_{1,2}$ decrease
and/or the electron dissipation rate $1/\tau_{\mathrm{NV}}$ increases, thus
our analytical results begin to agree with the exact numerical results. (iii)
For successively small $R$, the analytical dissipation rates $1/T_{1,2}$
(solid lines) tend to diverge, while the numerical results (dashed lines)
exhibit a maximum value $\sim1/\tau_{\mathrm{NV}}$. This is because at
sufficiently small $R$, the time scale of the NV dissipation $\tau
_{\mathrm{NV}}\sim1/R$ is longer than the nuclear spin dissipation and the NV
center becomes a non-Markovian bath. In this case, the electron-induced
nuclear spin dissipation rates $1/T_{1}$ and $1/T_{2}$ are upper limited by
the electron dissipation rate $\sim1/\tau_{\mathrm{NV}}$.

In Figs. \ref{ana_numer}(c) and \ref{ana_numer}(d), both the exact numerical
results (dashed lines) and our analytical formula (solid lines) show that the
nuclear spin dissipation rates $1/T_{1,2}$ increase rapidly with increasing
leakage rate $\gamma_{s2}$ from $m=0$ to $m=\pm1$ subspaces, due to the rapid
increase of the population $P_{-1_{e}}$ [see Eqs. (\ref{pure_pm1}) and
(\ref{relax_pm1})]. When $\gamma_{s2}=0$, the population $P_{-1_{e}}=0$, so
our analytical formula gives vanishing nuclear spin dissipation rates, while
the exact numerical results give a extremely small dissipation rates. This
residue dissipation comes from the process involving the electron spin flip,
which have been neglected in our theory since it is strongly suppressed by the
large electron-nuclear energy mismatch away from the ground state and excited
state anticrossing. Nevertheness, for extremely small $\gamma_{s2}$
($=\gamma_{s1}/1000$), it is responsible for the small difference between the
analytical results (blue solid line) and the exact numerical results (blue
dashed line) at large optical pumping rate in Fig. \ref{ana_numer}(d).

\begin{figure}[ptb]
\includegraphics[width=\columnwidth]{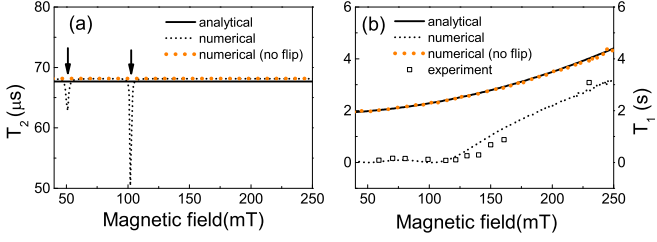}\caption{Magnetic field
dependence of the nuclear spin studied by Dreau \textit{et al.}%
\cite{DreauPRL2013}: (a) $T_{1}$ and (b) $T_{2}$ times from our analytical
formula (solid lines) and numerical simulations including (black dashed lines)
or excluding (orange dashed lines) the electron spin-flip terms in
$\mathbf{\hat{F}}$. The experimental results (empty squares) is also shown for
comparison. The two arrows indicate the ground state and excited state
anticrossings.}%
\label{exp_theory}
\end{figure}

Next we study the magnetic field dependence of $T_{1}$ and $T_{2}$ and compare
them with the experimental measurements \cite{DreauPRL2013}. For the nuclear
spin dephasing time $T_{2}$ in Fig. \ref{exp_theory}(a), away from the ground
state and excited anticrossing of the NV center [indicated by arrows in Fig.
\ref{exp_theory}(a)], our analytical formula agree well with the exact
numerical results, whether or not the electron spin flip terms in
$\mathbf{\hat{F}}$ is included. This indicates that the contribution involving
the electron spin flip is negligibly small compared with the contribution not
involving the electron spin flip.

In deriving Eq. (\ref{relax_pm1}) for the nuclear spin relaxation rates, we
have neglected the small nuclear spin level splitting $|\gamma_{N}\mathbf{B}%
|$. When this effect is included, the analytical expressions for $\Gamma_{\pm
}$ are given in Eq. (\ref{GAMMA_APPEND}), which shows a Lorentzian dependence
on the magnetic field $\Gamma_{\pm}\propto1/(|\mathbf{B}|^{2}+\delta
_{\mathrm{B}}^{2})$ with a characteristic width
\end{subequations}
\[
\delta_{\mathrm{B}}=\sqrt{\frac{R^{2}}{(2R+\gamma_{1})^{2}+2(R+\gamma
_{1})\gamma_{s1}+\gamma_{s1}^{2}}}\frac{\gamma_{s1}}{|\gamma_{N}|}.
\]
Under saturated pumping, as is usually used for optical readout, this width
$\sim\gamma_{1}/|\gamma_{N}|\sim500$ mT. By contrast, the contribution
involving the electron spin flip also has a Lorentzian dependence on the
magnetic field, but with a much smaller characteristic width $\sim\gamma
_{1}/\gamma_{e}\sim1$ $\mathrm{mT}$. The magnetic dependence of the relaxation
time of $^{13}\mathrm{C}$ nucleus has been measured by Dreau \textit{et al.
}\cite{DreauPRL2013}. They found that the anisotropic components
$A_{g,zx}=A_{g,xz}$ and $A_{g,zy}=A_{g,yz}$ of the ground state HFI
significantly contribute to the nuclear spin relaxation. The compoment
$A_{g,z,z}$ has been measured to be $0.25$ $\mathrm{MHz}$, while the other
components are not clear. Here we assume $\mathbf{A}_{e}=\mathbf{A}_{g}$ with
an isotropic diagonal components $A_{g,x,x}=A_{g,y,y}=A_{g,z,z}=0.25$ MHz and
a small anisotropic component $A_{g,z,x}=A_{g,x,z}=1.5$ kHz and $A_{g,zy}%
=A_{g,yz}=0$. Figure \ref{exp_theory}(b) shows that the exact numerical
results obtained by directly solving Eq. (\ref{ME0}) agree well with the
experimentally measured $T_{1}$ time \cite{DreauPRL2013}. As discussed
previously, the exact numerical results contain two contributions: the one not
involving the electron spin flip (which is treated by our analytical formula)
and the one involving the electron spin flip (which is not treated by our
analytical formula). Figure \ref{exp_theory}(b) shows that our analytical
formula provides an accurate description to the former contribution, although
in the present case the latter contribution dominates because of the much
larger isotropic HFI $\sim0.25$ MHz compared with the anisotropic HFI $\sim1$
kHz. Finally, for the nuclear spin at lattice $\mathrm{O}$ as reported in the
supplement of Ref. \onlinecite{DreauPRL2013}, it has a much shorter relaxation
time $\sim$ $40\ \mathrm{ms}$ at $200$ $\mathrm{mT}$. Such short
relaxation time is obviously dominated by the mechanism of Eq.(\ref{relax_pm1})
,from which, we can estimate the anisotropic HFI component of this nuclear
spin to be $\sim20\ \mathrm{kHz}$.

\subsection{Magnetic field perpendicular to N-V axis}

\begin{figure}[ptb]
\includegraphics[width=\columnwidth]{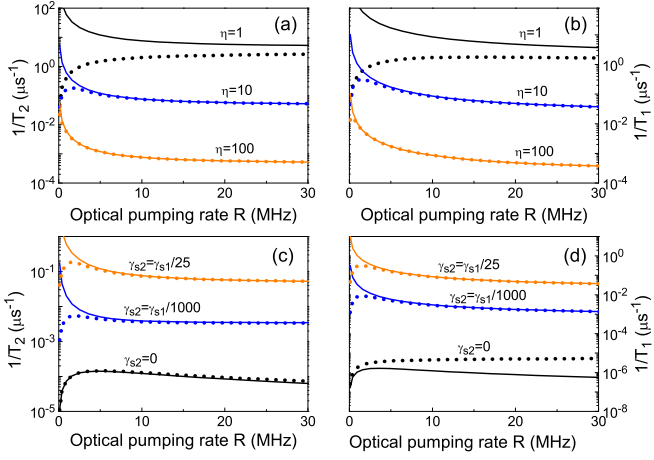}\caption{Comparison
of analytical (solid lines) and exact numerical results (dashed lines) for
nuclear spin $1/T_{2}$ [(a), (c)] and $1/T_{1}$ [(b), (d)] in a magnetic field
$B_{y}=10$ mT along the $y$ axis as functions of the optical pumping rate $R$.
Relevant parameters are $\mathbf{A}_{g/e}=\mathbf{A}_{g/e}(^{13}\mathrm{C}%
_{b})/\eta$ ($\eta=1,$ 10, 100), and $\gamma_{s2}=\gamma_{s1}/25$ in (a) and
(b); $\mathbf{A}_{g/e}=\mathbf{A}_{g/e}(^{13}\mathrm{C}_{b})/10$ and
$\gamma_{s2}=0,\gamma_{s1}/1000,\gamma_{s1}/25$ in (c) and (d).}%
\label{ana_num_trans}%
\end{figure}

Without loosing generality, we consider the magnetic field $\mathbf{B}%
=B_{y}\mathbf{e}_{y}$ along the $y$ axis of the conventional coordinate. In
this case, the precession frequencies $\mathbf{a}_{g}=-(2\gamma_{e}%
B_{y}/D_{\mathrm{gs}})\mathbf{e}_{y}\cdot\mathbf{A}_{g}$ and $\mathbf{a}%
_{e}=-(2\gamma_{e}B_{y}/D_{\mathrm{es}})\mathbf{e}_{y}\cdot\mathbf{A}_{e}$ are
proportional to the magntic field. The nuclear spin precession frequency
$\boldsymbol{\bar{\omega}}\equiv\gamma_{N}\mathbf{B}+P_{0_{g}}\mathbf{a}%
_{g}+P_{0_{e}}\mathbf{a}_{e}$ deviates from the $z$ axis. In this case, both
$\Gamma_{\varphi}^{(1)},\Gamma_{\pm}^{(1)}$ [see Eqs. (\ref{pure_pm1}) and
(\ref{relax_pm1})] from the $m=\pm1$ subspace and $\Gamma_{\varphi}%
^{(0)},\Gamma_{\pm}^{(0)}$ from the $m=0$ subspace are nonzero. For
$\Gamma_{\varphi}^{(1)}$ and $\Gamma_{\pm}^{(1)}$, the quantities
$b_{g,Z},\mathbf{b}_{g,\perp}$, etc. are defined in the tilted coordinate
$\mathbf{e}_{X},\mathbf{e}_{Y},\mathbf{e}_{Z}\equiv\boldsymbol{\bar{\omega}%
}/|\boldsymbol{\bar{\omega}}|$ that differs from the conventional coordinate
$(\mathbf{e}_{x},\mathbf{e}_{y},\mathbf{e}_{z})$.

First, we compare our analytical formula for the nuclear spin $1/T_{1}$ and
$1/T_{2}$ to the exact numerical results from directly solving the
electron-nuclear coupled equations of motion [Eq. (\ref{ME0})]. To see how our
analytical fomula becomes progressively applicable when going from the
non-Markovian regime to the Markovian regime, we start from the strongly
coupled nuclear spin\ $^{13}\mathrm{C}_{\mathrm{b}}$ and downscale its HFI
tensors $\mathbf{A}_{g}(^{13}\mathrm{C}_{b})$ and $\mathbf{A}_{e}%
(^{13}\mathrm{C}_{b})$ [see Eqs. (\ref{HFI})] by a factor $\eta=1,10,$ and
$100$ to decrease the nuclear spin dissipation. The nuclear spin $1/T_{2}$ and
$1/T_{1}$ shown in Fig. \ref{ana_num_trans} show very similar behaviors to the
case when the magnetic field is along the N-V axis [cf. Fig. (\ref{ana_numer}%
)], including the saturation at large optical pumping rate $R$ and the
improved agreement between the analytical results and the numerical results
with increasing $\eta$ and/or $R$. In particular, Figs. \ref{ana_num_trans}(c)
and \ref{ana_num_trans}(d) show that $1/T_{1,2}$ increase rapidly with the
leakage rate $\gamma_{s2}$, indicating that in addition to the contributions
$\Gamma_{\varphi}^{(1)}$ and $\Gamma_{\pm}^{(1)}$ from $m=\pm1$ subspace, the
contributions $\Gamma_{\varphi}^{(0)}$ and $\Gamma_{\pm}^{(0)}$ from the $m=0$
subspace also increase with $\gamma_{s2}$. For $\gamma_{s2}=0$, the nuclear
spin dissipation becomes very slow. In this case, the contribution from the
processes involving the electron spin flip (not included in our analytical
treatment) is no longer negligible. This leads to the discrepancy between the
analytical results (black solid lines) and the numerical results (black dotted
lines) in Fig. \ref{ana_num_trans}(d).

\begin{figure}[ptb]
\includegraphics[width=\columnwidth]{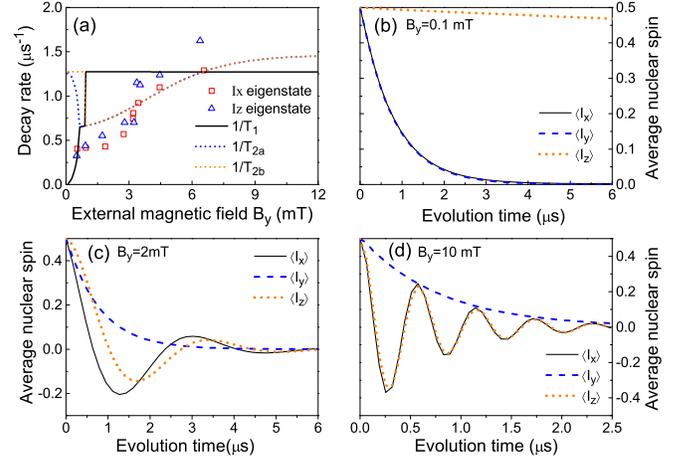}\caption{(a) Nuclear spin
$1/T_{1}$ (solid line) and $1/T_{2}$ (dashed lines) from numerically solving
Eq. (\ref{ME0}) compared with the experimentally measured decay for the
initial state being an eigenstate of $\hat{I}_{x}$ (squares) and $\hat{I}_{z}$
(triangles). (b)-(d) show the dissipative evolution under (b) $B_{y}%
=0.1\ \mathrm{mT}$, (c) $B_{y}=2\ \mathrm{mT,}$ and (d) $B_{y}=10\ \mathrm{mT}%
$ for the initial state being an eigenstate of $\hat{I}_{x}$ (solid line),
$\hat{I}_{y}$ (dashed line), and $\hat{I}_{z}$ (dotted line). The parameters
are $R=6.4$ MHz and $\gamma_{s2}=\gamma_{s1}/30$.}%
\label{decayrate}%
\end{figure}

Finally we set the scale factor $\eta=1$ and compare our theoretical results
with the experimental measurements \cite{DuttScience2007}. In this case, the
strong HFI makes the NV center a highly non-Markovian bath, so our analytical
theory only provides a qualitative description for the nuclear spin
dissipation. Since $\mathbf{A}_{g}(^{13}\mathrm{C}_{b})$ and $\mathbf{A}%
_{e}(^{13}\mathrm{C}_{b})$ [see Eqs. (\ref{HFI})] are approximately isotropic,
$\mathbf{a}_{g}$ and $\mathbf{a}_{e}$ are almost along the $y$ axis, while
$\mathbf{b}_{g}$ and $\mathbf{b}_{e}$ are approximately along the
$\mathbf{e}_{z}$ axis. For relatively large $B_{y}$, the magnetic field term
$\gamma_{N}B_{y}\mathbf{e}_{y}$ and the HFI contribution $\mathbf{a}%
_{g},\mathbf{a}_{e}\propto B_{y}\mathbf{e}_{y}$ dominates the average nuclear
spin precession frequency $\boldsymbol{\bar{\omega}}$, so the nuclear spin
quantization axis $\mathbf{e}_{Z}\propto\boldsymbol{\bar{\omega}}$ is almost
along the $y$ axis. Since $\mathbf{a}_{g}$ and $\mathbf{a}_{e}$ ($\mathbf{b}%
_{g}$ and $\mathbf{b}_{e}$) are nearly parallel (perpendicular) to
$\mathbf{e}_{Z}$, the $m=0$ $(m=\pm1$) subspace mainly contribute to the
nuclear spin pure dephasing (relaxation), so that $\Gamma_{\varphi}%
\approx\Gamma_{\varphi}^{(0)}$ increase quadratically with the magnetic field,
while $\Gamma_{\pm}\approx\Gamma_{\pm}^{(1)}$ is nearly independent of the
magnetic field. In other words, we expect that the nuclear spin $1/T_{2}$ to
increase appreciably with $B_{y}$ and the nuclear spin $1/T_{1}$ to be nearly
independent of $B_{y}$, as confirmed in Fig. \ref{decayrate}(a). According to
the Bloch equation Eq. (\ref{EOM_AVEI}), since the experimentally used initial
states are eigenstates of $\hat{I}_{z}$ and $\hat{I}_{x}$, their decay time is
largely determined by $T_{2}$. Indeed, for $B_{y}\gg1$ mT, Fig.
\ref{decayrate}(a) shows reasonable agreement between the numerically
calculated $1/T_{2}$ and the experimentally measured decay time of different
initial states. Note that the two-fold degenerate $1/T_{2}$ correspond to
identical decay of $\langle\hat{I}_{x}\rangle$ and $\langle\hat{I}_{z}\rangle$
[see Fig. \ref{decayrate}(d)]. By contrast, for $B_{y}\rightarrow0$, the
average nuclear spin precession frequency $\boldsymbol{\bar{\omega}}$ is
dominated by a small term $\langle\hat{S}_{g,z}\rangle\mathbf{b}_{g}%
+\langle\hat{S}_{e,z}\rangle\mathbf{b}_{e}$ along the $z$ axis (neglected in
our analytical treatment). In this case, the fluctuation of $\mathbf{b}_{g}$
and $\mathbf{b}_{e}$ of the $m=\pm1$ subspace mainly contribute to nuclear
spin pure dephasing, while the fluctuation of $\mathbf{a}_{g}$ and
$\mathbf{a}_{e}$ of the $m=0$ subspace mainly contribute to nuclear spin
relaxation. Correspondingly, in Fig. \ref{decayrate}(a), the nuclear spin
relaxation $1/T_{1}\propto B_{y}^{2}$ vanishes at $B_{y}=0$, while the
nuclear spin $1/T_{2}$ is two-fold degenerate, corresponding to near identical
decay of $\langle\hat{I}_{x}\rangle$ and $\langle\hat{I}_{y}\rangle$ [see Fig.
\ref{decayrate}(a)]. Due to the switch of the nuclear spin quantization axis
at intermediate magnetic field $B_{y}\sim1$ mT, the association of the solid
line with $1/T_{1}$ and the dashed lines with $1/T_{2}$ in Fig.
\ref{decayrate}(a) near the crossover region is meaningless.

\section{Conclusion}

We have presented a numerical and analytical study for the nuclear spin
dephasing and relaxation induced by an optically illuminated NV center at room
temperature. When the NV center undergoes a single cyclic transitions, our
analytical results provide a physically transparent interpretation that
substantiates the previous results \cite{JiangPRL2008} and demonstrate the
possibility to control the nuclear spin dissipation by tuning the magnetic
field \cite{JiangPRL2008}. For general optical illumination of the NV center
incorporating finite inter-system crossing, our numerical results agree with
the experimental measurements \cite{DreauPRL2013}. Our analytical results
suggests that the random hopping between the $m=0$ (or $m=\pm1$) triplet
states and the corresponding remained subspace of the NV center could
significantly contribute to nuclear spin dissipation. This means that
increasing the spin polarization degree of NV center would effectively
suppress the optical induced dissipation process. This contribution referred
here is not suppressed under saturated optical pumping and provides a possible
solution to the puzzling observation of nuclear spin dephasing in zero
magnetic field \cite{DuttScience2007}.

\appendix{}

\section{Lindblad master equation of nuclear spin}

Here we derive a closed equation of motion for the nuclear spin from the
coupled equation of motion Eq. (\ref{ME0}), where $\mathcal{L}_{e}$ is the
Liouville superoperator of the two-level or seven-level NV model. First, we
calculate the steady state density matrix $\hat{P}$ of the NV center from
$\mathcal{L}_{e}\hat{P}=0$. For the two-level model, the steady state
populations on $|e\rangle$ and $|g\rangle$ are $P_{e}=R/(2R+\gamma_{1})$ and
$P_{g}=1-P_{e}$, where $R=2\pi(\Omega_{R}/2)^{2}\delta^{((\gamma_{1}%
+\gamma_{\varphi})/2)}(\Delta)$ is the optical transition rate from
$|g\rangle$ to $|e\rangle$ and $\delta^{(\gamma)}(x)=(\gamma/\pi
)/(x^{2}+\gamma^{2})$ is the broadened $\delta$-function. For the seven-level
model at room temperature, due to the large orbital dephasing rate
$\gamma_{\varphi}\sim10^{7}$ MHz, the spin-conserving optical transition rates
from the ground orbital $|g\rangle$ to the excited orbital $|e\rangle$ are all
equal to $R\approx\Omega_{R}^{2}/\gamma_{\varphi}$ for different spin states.
The steady-state population on $|0_{g}\rangle$ is%
\[
P_{0{}_{g}}\approx\frac{R+\gamma_{1}+2\gamma_{s2}}{2R+\gamma_{1}+2\gamma
_{s2}(\frac{2R+\gamma_{1}+2\gamma_{s1}}{\gamma_{s1}}+\frac{R}{\gamma_{s}})}.
\]
The populations on other NV levels are%
\begin{align*}
P_{0{}_{e}}  &  \approx\frac{R}{R+\gamma_{1}+2\gamma_{s2}}P_{0{}_{g}},\\
P_{\pm1_{e}}  &  \approx\frac{R}{R+\gamma_{1}+\gamma_{s1}}P_{\pm1_{g}}%
\approx\frac{\gamma_{s2}}{\gamma_{s1}}P_{0_{e}},
\end{align*}
and $P_{S}\approx(2\gamma_{s1}/\gamma_{s})P_{\pm1_{e}}$. When the leakage from
the $m=0$ subspace to the $m=\pm1$ subspaces are neglected by setting
$\gamma_{s2}=0$, we have $P_{\pm1_{e}}=P_{\pm1_{g}}=P_{S}=0$ and $P_{0_{e}%
}=1-P_{0_{g}}=R/(2R+\gamma_{1})$, which recovers the two-level fluctuator model.

Second, we decompose the HFI into the mean-field part $\langle\hat{\mathbf{F}%
}\rangle_{e}\cdot\hat{\mathbf{I}}$ and the fluctuation part $(\hat{\mathbf{F}%
}-\langle\hat{\mathbf{F}}\rangle_{e})\cdot\hat{\mathbf{I}}\equiv
\mathbf{\tilde{F}}\cdot\hat{\mathbf{I}}$, where $\langle\hat{\mathbf{F}%
}\rangle_{e}\equiv\operatorname*{Tr}\hat{\mathbf{F}}\hat{P}$ is the average
Knight field from the NV center, e.g., $\langle\hat{\mathbf{F}}\rangle
_{e}=P_{g}\boldsymbol{\omega}_{g}+P_{e}\boldsymbol{\omega}_{e}$ for the
two-level model and $\langle\hat{\mathbf{F}}\rangle_{e}=P_{0_{g}}%
\mathbf{a}_{g}+P_{0_{e}}\mathbf{a}_{e}$ for the seven-level model. Under this
decomposition, Eq. (\ref{ME0}) becomes%
\begin{equation}
\dot{\rho}=\mathcal{L}_{e}\hat{\rho}-i[\boldsymbol{\bar{\omega}}\cdot
\hat{\mathbf{I}},\hat{\rho}]-i[\mathbf{\tilde{F}}\cdot\hat{\mathbf{I}}%
,\hat{\rho}],
\end{equation}
where $\boldsymbol{\bar{\omega}}\equiv\gamma_{N}\mathbf{B}+\langle
\hat{\mathbf{F}}\rangle_{e}$ is the total magnetic field that defines the
nuclear spin quantization axis. Consequently, the nuclear spin dephasing and
relaxation should be defined in the cartesian frame $(\mathbf{e}%
_{X},\mathbf{e}_{Y},\mathbf{e}_{Z})$, where $\mathbf{e}_{Z}\equiv
\boldsymbol{\bar{\omega}}/|\boldsymbol{\bar{\omega}}|$.

Third, we decompose $\mathbf{\tilde{F}}\cdot\hat{\mathbf{I}}$ into the sum of
the longitudional part $\tilde{F}_{Z}\hat{I}_{Z}$ and the transverse part
$(\tilde{F}_{+}\hat{I}_{-}+\tilde{F}_{-}\hat{I}_{+})/2$, where $O_{\pm}\equiv
O_{X}\pm iO_{Y}$. Then treating $\mathbf{\tilde{F}}\cdot\hat{\mathbf{I}}$ by
the adiabatic approximation \cite{WangArxiv2015} up to the second order gives
Eq. (\ref{ME_N}) for the nuclear spin density matrix $\hat{p}(t)\equiv
\operatorname*{Tr}_{e}\hat{\rho}(t)$, where
\[
\Gamma_{\varphi}=\operatorname{Re}\int_{0}^{+\infty}\mathrm{Tr}_{e}\tilde
{F}_{Z}(e^{\mathcal{L}_{e}t}\tilde{F}_{Z}\hat{P})dt\equiv\operatorname{Re}%
\langle\hat{F}_{Z};\hat{F}_{Z}\rangle_{0}%
\]
is the nuclear spin pure dephasing rate due to the fluctuation of $\hat{F}%
_{Z}$ at zero frequency, and
\[
\Gamma_{\pm}=\frac{1}{2}\operatorname{Re}\int_{0}^{+\infty}\mathrm{Tr}%
_{e}\tilde{F}_{\pm}(e^{(\mathcal{L}_{e}\mp i|\boldsymbol{\bar{\omega}}%
|)t}\tilde{F}_{\mp}\hat{P})dt\equiv\frac{1}{2}\operatorname{Re}\langle\hat
{F}_{\pm};\hat{F}_{\mp}\rangle_{\pm|\boldsymbol{\bar{\omega}}|},
\]
is the nuclear spin-flip rates due to the fluctuation of $\hat{F}_{\mp}$ at
the nuclear spin precession frequency $|\boldsymbol{\bar{\omega}}|$, and
\[
\langle\hat{a};\hat{b}\rangle_{\omega}\equiv\int_{0}^{+\infty}\mathrm{Tr}%
_{e}\tilde{a}e^{(\mathcal{L}_{e}-i\omega)t}\tilde{b}\hat{P}dt=-\mathrm{Tr}%
_{e}\tilde{a}(\mathcal{L}_{e}-i\omega)^{-1}\tilde{b}\hat{P}%
\]
is the steady-state correlation at frequency $\omega$ between the fluctuation
$\tilde{a}\equiv\hat{a}-\operatorname*{Tr}\hat{a}\hat{P}$ and the fluctuation
$\tilde{b}\equiv\hat{b}-\operatorname*{Tr}\hat{b}\hat{P}$.

For the seven-level NV model, we have
\begin{align*}
\tilde{F}_{Z}  &  =b_{g,Z}\tilde{S}_{g,z}+b_{e,Z}\tilde{S}_{e,z}+a_{g,Z}%
\tilde{\sigma}_{0_{g},0_{g}}+a_{e,Z}\tilde{\sigma}_{0_{e},0_{e}},\\
\tilde{F}_{\pm}  &  =b_{g,\pm}\tilde{S}_{g,z}+b_{e,\pm}\tilde{S}%
_{e,z}+a_{g,\pm}\tilde{\sigma}_{0_{g},0_{g}}+a_{e,\pm}\tilde{\sigma}%
_{0_{e},0_{e}},
\end{align*}
where $\tilde{O}\equiv O-\operatorname*{Tr}\hat{O}\hat{P}$ is the fluctuation
part of electron operator $\hat{O}$, $a_{g,\pm}\equiv a_{g,X}\pm ia_{g,Y}$ and
$b_{g,\pm}\equiv b_{g,X}\pm ib_{g,Y}$, etc. We can verify that the group
$\tilde{S}_{e,z},\tilde{S}_{g,z}$ and the group $\tilde{\sigma}_{0_{e},0_{e}%
},\tilde{\sigma}_{0_{g},0_{g}}$ have vanishing cross-correlation, so
$\Gamma_{\pm}=\Gamma_{\pm}^{(1)}+\Gamma_{\pm}^{(0)}$ and $\Gamma_{\varphi
}=\Gamma_{\varphi}^{(1)}+\Gamma_{\varphi}^{(0)}$ can be written as the sum of
the contributions from the $m=\pm1$ subspaces:
\begin{align*}
\Gamma_{\varphi}^{(1)}  &  =\operatorname{Re}(|b_{g,Z}|^{2}\langle\hat
{S}_{g,z};\hat{S}_{g,z}\rangle_{0}+|b_{e,Z}|^{2}\langle\hat{S}_{e,z};\hat
{S}_{e,z}\rangle_{0})\\
&  \quad+\operatorname{Re}b_{g,Z}b_{e,Z}\left(  \langle\hat{S}_{g,z};\hat
{S}_{e,z}\rangle_{0}+\langle\hat{S}_{e,z};\hat{S}_{g,z}\rangle_{0}\right)  ,\\
\Gamma_{\pm}^{(1)}  &  \approx\frac{1}{2}\operatorname{Re}(|\mathbf{b}%
_{g,\perp}|^{2}\langle\hat{S}_{g,z};\hat{S}_{g,z}\rangle_{\pm|\boldsymbol{\bar
{\omega}}|}+|\mathbf{b}_{e,\perp}|^{2}\langle\hat{S}_{e,z};\hat{S}%
_{e,z}\rangle_{\pm|\boldsymbol{\bar{\omega}}|})\\
&  \quad+\frac{1}{2}\operatorname{Re}(b_{g,\pm}b_{e,\mp}\langle\hat{S}%
_{g,z};\hat{S}_{e,z}\rangle_{\pm|\boldsymbol{\bar{\omega}}|}+b_{g,\mp}%
b_{e,\pm}\langle\hat{S}_{e,z};\hat{S}_{g,z}\rangle_{\pm|\boldsymbol{\bar
{\omega}}|})
\end{align*}
and the contributions from the $m=0$ subspaces:%
\begin{align*}
\Gamma_{\varphi}^{(0)}  &  =\operatorname{Re}\left(  a_{e,Z}^{2}\langle
\hat{\sigma}_{0_{e},0_{e}};\hat{\sigma}_{0_{e},0_{e}}\rangle_{0}+a_{g,Z}%
^{2}\langle\hat{\sigma}_{0_{g},0_{g}};\hat{\sigma}_{0_{g},0_{g}}\rangle
_{0}\right) \\
&  \quad+\operatorname{Re}a_{g,Z}a_{e,Z}\left(  \langle\hat{\sigma}%
_{0_{e},0_{e}};\hat{\sigma}_{0_{g},0_{g}}\rangle_{0}+\langle\hat{\sigma
}_{0_{g},0_{g}};\hat{\sigma}_{0_{e},0_{e}}\rangle_{0}\right)  ,\\
\Gamma_{\pm}^{(0)}  &  \approx\frac{1}{2}\operatorname{Re}(|\mathbf{a}%
_{e,\perp}|^{2}\langle\hat{\sigma}_{0_{e},0_{e}};\hat{\sigma}_{0_{e},0_{e}%
}\rangle_{\pm|\boldsymbol{\bar{\omega}}|}+|\mathbf{a}_{g,\perp}|^{2}%
\langle\hat{\sigma}_{0_{g},0_{g}};\hat{\sigma}_{0_{g},0_{g}}\rangle
_{\pm|\boldsymbol{\bar{\omega}}|})\\
&  \quad+\frac{1}{2}\operatorname{Re}(a_{g,\mp}a_{e,\pm}\langle\hat{\sigma
}_{0_{e},0_{e}};\hat{\sigma}_{0_{g},0_{g}}\rangle_{\pm|\boldsymbol{\bar
{\omega}}|}+a_{g,\pm}a_{e,\mp}\langle\hat{\sigma}_{0_{g},0_{g}};\hat{\sigma
}_{0_{e},0_{e}}\rangle_{\pm|\boldsymbol{\bar{\omega}}|}).
\end{align*}
Using the analytical expressions for the above correlation functions in
Appendix B, we obtain $\Gamma_{\varphi}^{(1)}$ as Eq. (\ref{pure_pm1}) and
\begin{align}
\Gamma_{\pm}^{(1)}  &  =\frac{P_{-1_{e}}}{\gamma_{s1}}f(|\boldsymbol{\bar
{\omega}}|)|\mathbf{b}_{e,\perp}|^{2}\left(  1+\frac{R+\gamma_{1}+\gamma_{s1}%
}{R}\frac{|\boldsymbol{\bar{\omega}}|^{2}}{R\gamma_{s1}}\right)
\label{GAMMA_APPEND}\\
&  \quad+\frac{P_{-1_{e}}}{\gamma_{s1}}f(|\boldsymbol{\bar{\omega}%
}|)|\mathbf{b}_{g,\perp}|^{2}\frac{R+\gamma_{1}+\gamma_{s1}}{R}\left(
\frac{R+\gamma_{1}+\gamma_{s1}}{R}+\frac{|\boldsymbol{\bar{\omega}}|^{2}%
}{R\gamma_{s1}}\right) \nonumber\\
&  \quad+\frac{P_{-1_{e}}}{\gamma_{s1}}f(|\boldsymbol{\bar{\omega}%
}|)(\mathbf{b}_{g,\perp}\cdot\mathbf{b}_{e,\perp})\frac{2R+2\gamma_{1}%
+\gamma_{s1}}{R}\left(  1-\frac{|\boldsymbol{\bar{\omega}}|^{2}}{R\gamma_{s}%
}\right) \nonumber\\
&  \quad+\frac{P_{-1_{e}}}{\gamma_{s1}}f(|\boldsymbol{\bar{\omega}%
}|)(\mathbf{b}_{g}\times\mathbf{b}_{e})_{Z}\frac{|\boldsymbol{\bar{\omega}}%
|}{R}\frac{2R+\gamma_{1}+\gamma_{s1}}{R},\nonumber
\end{align}
where
\[
f(\omega)=\frac{R^{2}\gamma_{s1}^{2}}{R^{2}\gamma_{s1}^{2}+[(2R+\gamma
_{1})^{2}+2(R+\gamma_{1})\gamma_{s1}+\gamma_{s1}^{2}]\omega^{2}+\omega^{4}}.
\]

\section{Steady-state correlation functions}

Here we use the equation of motion method to evaluate the eight correlation
functions. For example, the correlation function $\langle\hat{\sigma}%
_{0_{e},0_{e}};\hat{\sigma}_{0_{e},0_{e}}\rangle_{0}$ can be written as
$-\operatorname*{Tr}_{e}\tilde{\sigma}_{0_{e},0_{e}}\hat{X}=-X_{0_{e},0_{e}}$,
where $\hat{X}\equiv\mathcal{L}_{e}^{-1}\tilde{\sigma}_{0_{e},0_{e}}\hat{P}$
obeys $\mathrm{Tr}_{e}\hat{X}=0$ and $X_{ij}\equiv\langle i|\hat{X}|j\rangle$.
The large orbital dephasing rate $\gamma_{\varphi}\sim10^{7}$ MHz allows us to
neglect the off-diagonal coherence of the electron and only keep the diagonal
populations $P_{i}\equiv\langle i|\hat{P}|i\rangle$. The equations of motion
of $X_{ij}$ is obtained by taking the $(i,j)$ matrix element of $\mathcal{L}%
_{e}\hat{X}=\tilde{\sigma}_{0_{e},0_{e}}\hat{P}$. We find that the equations
of motion of the diagonal (off-diagonal) elements of $\hat{X}$ involves the
off-diagonal (diagonal) elements. By eliminating the off-diagonal elements in
favor of the diagonal elements, we obtain
\begin{align}
-(\gamma_{1}+2\gamma_{s2}+R)X_{0_{e},0_{e}}+RX_{0_{g},0_{g}}  &  =P_{0_{e}%
}(1-P_{0_{e}}),\nonumber\\
\gamma_{s}X_{S,S}+(\gamma_{1}+R)X_{0_{e},0_{e}}-RX_{0_{g},0_{g}}  &
=-P_{0_{e}}P_{0_{g}},\nonumber\\
-\gamma_{s}X_{S,S}+\gamma_{s1}(X_{-1_{e},-1_{e}}+X_{+1_{e},+1_{e}})  &
=-P_{S}P_{0_{g}},\nonumber\\
-(\gamma_{1}+\gamma_{s1}+R)X_{-1_{e},-1_{e}}+RX_{-1_{g},-1_{g}}  &
=-P_{-1_{e}}P_{0_{g}},\nonumber\\
-(\gamma_{1}+\gamma_{s1}+R)X_{+1_{e},+1_{e}}+RX_{+1_{g},+1_{g}}  &
=-P_{+1_{e}}P_{0_{g}},\nonumber\\
(\gamma_{1}+R)X_{-1_{e},-1_{e}}+\gamma_{s2}X_{0_{e},0_{e}}-RX_{-1_{g},-1_{g}}
&  =-P_{-1_{g}}P_{0_{e}},\nonumber\\
(\gamma_{1}+R)X_{+1_{e},+1_{e}}+\gamma_{s2}X_{0_{e},0_{e}}-RX_{+1_{g},+1_{g}}
&  =-P_{+1_{g}}P_{0_{e}},\nonumber
\end{align}
where $\Delta_{i,j}$ is the energy difference between the electron state
$|i\rangle$ and $|j\rangle$ in the rotating frame of the pumping laser.
Solving the above equations gives the correlation function $\langle\hat
{\sigma}_{0_{e},0_{e}};\hat{\sigma}_{0_{e},0_{e}}\rangle_{0}=-X_{0_{e},0_{e}}$
as\begin{widetext}%
\[
\langle\hat{\sigma}_{0_{e},0_{e}};\hat{\sigma}_{0_{e},0_{e}}\rangle_{0}%
=\frac{P_{0_{e}}P_{0_{g}}}{1+\eta}\frac{1}{2R+\gamma_{1}}+\frac{P_{0_{e}%
}(1-P_{0_{e}}-P_{0_{g}})}{1+\eta}\left(  \frac{R+\gamma_{s}}{2R+\gamma_{1}%
}\frac{1}{\gamma_{s}}+\frac{1}{\gamma_{s1}}\right)  +\frac{P_{0_{e}}%
(P_{-1_{g}}+P_{+1_{g}})}{1+\eta}\frac{1}{2R+\gamma_{1}}-\frac{P_{0_{e}}P_{S}%
}{1+\eta}\frac{1}{\gamma_{s1}}.
\]
where $\eta=2\gamma_{s2}/\gamma_{s1}+2(R+2\gamma_{s})\gamma_{s2}/(\gamma
_{s}(2R+\gamma_{1}))$ is a dimensionless constant much smaller than unity
since $\gamma_{s2}\ll\gamma_{s1},\gamma_{s}$. Using the same method, the other
correlation functions are obtained as:
\begin{align*}
\langle\hat{\sigma}_{0_{g},0_{g}}(t)\hat{\sigma}_{0_{g},0_{g}}(0)\rangle_{0}
&  =\frac{P_{0_{g}}P_{0_{e}}}{1+\eta}\left(  \frac{1-2\gamma_{s2}/R}%
{2R+\gamma_{1}}+\frac{\eta}{R}\right)  +\frac{P_{0_{g}}(1-P_{0_{e}}-P_{0_{g}%
})}{1+\eta}\left(  \frac{1}{R}+\frac{\gamma_{s1}}{\gamma_{s}(2R+\gamma_{1}%
)}\right)  \frac{R+\gamma_{1}+2\gamma_{s2}}{\gamma_{s1}}\\
&  +\frac{P_{0_{g}}(P_{-1_{e}}+P_{+1_{e}})}{1+\eta}\frac{1}{R}\frac
{R+\gamma_{1}+2\gamma_{s2}}{2R+\gamma_{1}}-\frac{P_{0_{g}}P_{S}}{1+\eta}%
\frac{1}{R}\frac{R+\gamma_{1}+2\gamma_{s2}}{\gamma_{s1}},\\
\langle\hat{\sigma}_{0_{e},0_{e}}(t)\hat{\sigma}_{0_{g},0_{g}}(0)\rangle_{0}
&  =-\frac{P_{0_{g}}P_{0_{e}}}{1+\eta}\frac{1}{2R+\gamma_{1}}+\frac{P_{0_{g}%
}(1-P_{0_{e}}-P_{0_{g}})}{1+\eta}\left(  \frac{1}{\gamma_{s1}}+\frac
{R}{2R+\gamma_{1}}\frac{1}{\gamma_{s}}\right)  +\frac{P_{0_{g}}(P_{-1_{g}%
}+P_{+1_{g}})}{(1+\eta)(2R+\gamma_{1})}-\frac{P_{0_{g}}P_{S}}{1+\eta}\frac
{1}{\gamma_{s1}},\\
\langle\hat{\sigma}_{0_{g},0_{g}}(t)\hat{\sigma}_{0_{e},0_{e}}(0)\rangle_{0}
&  =-\frac{P_{0_{e}}(1-P_{0_{e}})}{1+\eta}\frac{1+2\gamma_{s2}/R}%
{2R+\gamma_{1}}-2\frac{P_{0_{e}}P_{0_{g}}}{1+\eta}\left(  \frac{\gamma_{s2}%
}{R\gamma_{s1}}+\frac{\gamma_{s2}}{(2R+\gamma_{1})\gamma_{s}}\right)
+\frac{P_{0_{e}}(P_{-1_{g}}+P_{+1_{g}})}{1+\eta}\left(  \frac{1}{R}%
-\frac{1-2\gamma_{s1}/R}{2R+\gamma_{1}}\right)  \\
&  +\frac{P_{0_{e}}(1-P_{0_{e}}-P_{0_{g}})}{1+\eta}\left(  \frac{R+\gamma_{1}%
}{R\gamma_{s1}}+\frac{R+\gamma_{1}}{2R+\gamma_{1}}\frac{1}{\gamma_{s}}\right)
-\frac{P_{0_{e}}P_{S}}{1+\eta}\frac{R+\gamma_{1}+2\gamma_{s2}}{R\gamma_{s1}}.
\end{align*}
\end{widetext}Similarly, the correlation functions at finite frequency are
obtained as
\begin{align}
&  \langle\hat{S}_{e,z};\hat{S}_{e,z}\rangle_{\omega}=\frac{2(R+i\omega
)}{R\gamma_{s1}+i(\gamma_{1}+\gamma_{s1}+2R)\omega-\omega^{2}}P_{-1_{e}%
},\nonumber\\
&  \langle\hat{S}_{g,z};\hat{S}_{g,z}\rangle_{\omega}=\frac{2(R+\gamma
_{1}+\gamma_{s1}+i\omega)}{R\gamma_{s1}+i(2R+\gamma_{1}+\gamma_{s1}%
)\omega-\omega^{2}}P_{-1_{g}},\nonumber\\
&  \langle\hat{S}_{e,z};\hat{S}_{g,z}\rangle_{\omega}=\frac{2R}{R\gamma
_{s1}+i(2R+\gamma_{1}+\gamma_{s1})\omega-\omega^{2}}P_{-1_{g}},\nonumber\\
&  \langle\hat{S}_{g,z};\hat{S}_{e,z}\rangle_{\omega}=\frac{2(R+\gamma_{1}%
)}{R\gamma_{s1}+i(2R+\gamma_{1}+\gamma_{s1})\omega-\omega^{2}}P_{-1_{e}%
}.\nonumber
\end{align}

When the leakage from the $m=0$ subspace to the $m=\pm1$ subspace are
neglected by setting $\gamma_{s2}=0$ (and hence $\eta=0$), only the
populations $P_{0_{e}}$ and $P_{0_{g}}$ are nonzero, so only the first term in
the above expressions survives:%
\begin{align*}
\langle\hat{\sigma}_{0_{e},0_{e}};\hat{\sigma}_{0_{e},0_{e}}\rangle_{0}  &
=\langle\hat{\sigma}_{0_{g},0_{g}}(t)\hat{\sigma}_{0_{g},0_{g}}(0)\rangle
_{0}=-\langle\hat{\sigma}_{0_{g},0_{g}}(t)\hat{\sigma}_{0_{e},0_{e}}%
(0)\rangle_{0}\\
&  =-\langle\hat{\sigma}_{0_{e},0_{e}}(t)\hat{\sigma}_{0_{g},0_{g}}%
(0)\rangle_{0}=\frac{P_{0_{e}}P_{0_{g}}}{2R+\gamma_{1}},
\end{align*}
which give Eqs. (\ref{GAMMA_TLS}) of the main text. For saturated pumping, we
have
\begin{align*}
\langle\hat{\sigma}_{0_{g},0_{g}};\hat{\sigma}_{0_{e},0_{e}}\rangle_{0}  &
=\langle\hat{\sigma}_{0_{e},0_{e}};\hat{\sigma}_{0_{e},0_{e}}\rangle
_{0}=\langle\hat{\sigma}_{0_{g},0_{g}};\hat{\sigma}_{0_{e},0_{e}}\rangle_{0}\\
&  =\langle\hat{\sigma}_{0_{e},0_{e}};\hat{\sigma}_{0_{g},0_{g}}\rangle
_{0}=\frac{\tilde{\tau}_{0}^{2}}{8\tilde{T}}%
\end{align*}
and hence Eqs. (\ref{GAMMA_SAT}) of the main text.


\end{document}